\documentstyle[12pt,aasms4]{article}
\begin{document}

\begin{center}
\title{The Detection of \raisebox{.6ex}{13}CO and Other Apparent Abundance Anomalies in the Secondary Stars of Long-Period Cataclysmic Variables}

\author{Thomas E. Harrison\raisebox{.6ex}{1,2}, Heather L. Osborne\raisebox{.6ex}{2}}

\affil{Department of Astronomy New Mexico State University, Box 30001, MSC 4500, Las Cruces, NM 88003-8001}

\authoremail{ tharriso@nmsu.edu, hosborne@nmsu.edu}

\author{Steve B. Howell}

\affil{WIYN Observatory and National Optical Astronomy Observatories, 950 North Cherry Avenue, Tucson, AZ 85726}

\authoremail{howell@noao.edu}

{\it Accepted for publication in the Astronomical Journal on 18 February,
2004}
\end{center}

\begin{flushleft}
\raisebox{.6ex}{1}Visiting Astronomer, Cerro Tololo Inter-American
Observatory, National Optical Astronomy
Observatory, which is operated by the Association of Universities for Research
in Astronomy, Inc.
(AURA) under cooperative agreement with the National Science Foundation.

\raisebox{.6ex}{2}Visiting Astronomer at the Infrared Telescope Facility,
which is operated by the University of Hawaii
under contract from the National Aeronautics and Space Administration.

\end{flushleft}

\noindent
{\it Subject headings:} cataclysmic variablesstars: infrared spectrastars:
individual (V442 Centauri, SY
Cancri, RU Pegasi, CH Ursae Majoris, MU Centauri, TT Crateris, AC Cancri, EM
Cygni, V426
Ophiuchi, SS Cygni, BV Puppis, AH Herculis)

\abstract{ 
We present moderate resolution (R $\geq$ 1,800) infrared {\it K}-band
spectra of twelve long-period (P\raisebox{-.6ex}{orb} $\geq$ 6
hr) cataclysmic variables. We detect absorption lines from the photospheres of
the secondary stars
in every system, even though two of them were undergoing outbursts. We have
attempted to assign
spectral types to each of the secondary stars, and these classifications are
generally consistent with
previous determinations/estimates. We find evidence for abundance anomalies
that include
enhancements and/or deficits for all of the species commonly found in {\it
K}-band spectra of G- and K-type dwarfs. There is, however, only one common
abundance anomaly: extremely weak CO features.
Only two of the twelve objects appeared to have normal levels of CO
absorption. We interpret this
as evidence for low carbon abundances. In addition, we detect
\raisebox{.6ex}{13}CO absorption in four of the twelve
objects. Depleted levels of \raisebox{.6ex}{12}C and enhanced levels of
\raisebox{.6ex}{13}C indicate that material that has been
processed in the CNO-cycle is finding its way into the photospheres of CV
secondary stars. In systems
with luminous accretion disks, we find that the spectrum of the secondary star
is contaminated by a
source that flattens (reddens) the continuum. While free-free or classical
accretion disk spectra are
flatter than the blackbody-like spectra of G and K dwarfs, removal of such
contamination from the
{\it K}-band data results in spectra in which the absorption features become
too strong to be consistent
with those of G and K dwarfs.}

\section{Introduction}

Cataclysmic variables (CVs) are short-period binary systems
consisting of a white dwarf
primary that is accreting material via Roche-lobe overflow from a low mass,
late-type secondary star.
The commonly proposed evolutionary history for cataclysmic variables (CVs)
establishes that the vast
majority of CV secondary stars have undergone very little evolution during
their lifetime (see Howell,
Nelson, and Rappaport 2001, and references therein). The formation of a CV
from a wide binary
containing two main sequence stars is envisaged to have three main phases:
First, the orbital
separation of the wide binary of the pre-CV is rapidly shrunk in a common
envelope phase where the
secondary star orbits inside the red giant photosphere of the white dwarf
progenitor. The second phase
is a very long epoch where gravitational radiation, or a magnetically
constrained wind from the
secondary star (magnetic braking) extracts angular momentum from the binary,
resulting in the
eventual contact of the photosphere of the secondary star with its Roche lobe.
The final phase begins
once the secondary star contacts its Roche lobe, mass transfer to the white
dwarf is initiated, and all
of the phenomena associated with CVs is observed. During the lifetime of the
mass transfer phase,
the overall mass of the secondary star is gradually reduced. Much of the
material accreted by the white
dwarf is believed to be lost from the typical CV system through numerous
classical novae eruptions.
In both the common envelope phases and during classical novae eruptions,
material with a peculiar
composition can be deposited in the photospheres of CV secondary stars.

Evidence for the existence of peculiar abundance patterns in CVs
is growing. For example,
using UV spectroscopy, Cheng et al. (1997) found that the carbon abundance was
5$\times$ solar, nitrogen
was 3$\times$ solar, and silicon was $\leq$ 0.1$\times$ solar for the white
dwarf in WZ Sge. They suggested that this
material was probably transferred from the secondary star. For VW Hyi, Sion et
al. (1997) found that
the white dwarf appeared to be deficient in carbon (0.5$\times$ solar), iron
(0.5$\times$ solar), and silicon (0.1$\times$
solar), but had an excess of nitrogen (5$\times$ solar), oxygen (2$\times$
solar), and phosphorous (900$\times$ solar).
Sion et al. (2001) found subsolar abundances for carbon (0.05$\times$ solar)
and silicon (0.1$\times$ solar) for the
white dwarf in RX And. In both CN Ori (Urban et al. 2000) and AH Her (Lyons et
al. 2001) subsolar
silicon abundances were found. Meanwhile, Sion et al. (1998) and Long \&
Gilliland (1999), estimated
that the carbon abundance of the white dwarf in U Gem is about 0.1$\times$
solar, while the nitrogen
abundance is about 4$\times$ solar. Harrison et al. (2000) found from infrared
spectroscopy that the
secondary star in U Gem appeared to have extremely weak CO features,
suggesting it was deficient
in either carbon, or oxygen. If so, the deficit of carbon on the white dwarf
for U Gem could be easily
explained by the transfer of carbon-poor material from the secondary star. In
addition, Harrison et al.
found that the secondary star of SS Cyg displayed weaker CO features than it
should for its spectral
type, along with an apparent magnesium deficit. Mennickent \& Diaz (2002)
report weak CO
absorption for the late-type secondary in VY Aqr. It appears that the both the
primary and secondary
stars in CVs have peculiar compositions.

Recently, G\"ansicke et al. (2003) have reported on anomalous N
V/C IV line flux ratios in new
{\it HST} STIS ultraviolet spectroscopy for several CVs, and compile a list of
ten CVs that all show
similar spectra. They conclude that these represent true abundance anomalies,
and that nitrogen is
strongly enhanced relative to carbon.

What could be the origin of such abundance anomalies? Sion (1999)
suggests the abundance
anomalies in the white dwarf photospheres may result from the nuclear
processing in classical novae
explosions. As shown by Jos\'e, Coc and Hernanz (2001), however, only
classical novae with very
massive white dwarfs (1.35 M$_{\sun}$) can produce nuclei such as
silicon or phosphorous in
thermonuclear burning (and such burning increases, not decreases the abundance
of silicon). White
dwarf masses in typical CV systems cluster near 0.6 M$_{\sun}$. The
presence of similar abundance patterns
in the secondary stars suggest that for most CVs, they are a more likely
source for material of peculiar
composition. Marks \& Sarna (1998) performed a detailed theoretical study of
the possible effects on
the surface abundances of the secondary star due to both evolutionary effects
in the secondary itself,
as well as that due to sweeping-up of CE material or matter accreted from
classical novae ejecta. All
such events could place thermonuclear processed material into the photosphere
of the secondary star.
Considering the first case, Marks \& Sarna find that the photospheric
chemistry of the secondary star
could show large abundance variations in carbon, nitrogen and oxygen from
evolutionary effects
alone. In this scenario, the CNO tricycle is operating in the secondary star
either before or during the
contact phase. As material is removed from the secondary star, layers where
the CNO tricycle was
operating are exposed, or mixed to the surface, creating abundance and
isotopic variations in CNO
species. Especially relevant is their predictions for an overall deficit in
carbon, enhancements in
nitrogen, and a dramatic change in the ratio of
\raisebox{.6ex}{12}C/\raisebox{.6ex}{13}C. For this work, however, Marks \&
Sarna only
considered initially massive secondary stars (1.0 - 1.5 M$_{\sun}$),
whereas Howell et al. (2001) have shown
that massive secondaries are likely to be present in only a small fraction of
CVs. 

Marks \& Sarna (1998) also performed a study of the effects on the
surface abundances of the
secondary star due to sweeping-up of CE material or by accreting classical
novae ejecta. They
concluded that any material acquired during the CE phase would be thoroughly
mixed into the
secondary star during the extended period between the CE phase and the time
the secondary contacted
its Roche lobe. Marks \& Sarna did find that if the process of accreting novae
ejecta was efficient,
dramatic abundance and isotopic anomalies could be present in the photospheres
of CV secondaries. 

G\"ansicke et al. (2003) suggest that the anomalous nitrogen to
carbon abundances are a natural
consequence of a scenario where the initial mass of the current donor star was
greater than that of the
white dwarf. As described by Schenker et al. (2002), in this situation a
short-lived phase of very high,
and dynamically unstable mass transfer quickly whittles away the outer layers
of the donor star,
leading to the production of a CV with a relatively normal mass ratio, but one
where the donor star
is now the CNO processed core of the massive donor. Schenker et al. propose
that the unusual CV
system AE Aqr has just completed this phase of evolution. As shown in their
Fig. 9, the surface
chemical abundance ratios of \raisebox{.6ex}{12}C/\raisebox{.6ex}{13}C, and
C/N drop by one or two orders of magnitude as the system
like AE Aqr evolves to become a CV. This scenario predicts depleted levels of
carbon, enhanced
levels of nitrogen and \raisebox{.6ex}{13}C, and that the donor stars will
have spectral types that are too late for their
orbital periods.

Thus, the detection and measurement of abundance anomalies in a
secondary star may provide
direct insight into the evolutionary history of a CV. Of course, it remains
quite possible that any
observed abundance anomalies might arise due to unusual excitation conditions
within the non-equilibrium photospheres of irradiated, mass-losing secondary
stars. We present new infrared spectra
of a dozen long period cataclysmic variables to search for additional
abundance anomalies in their
secondary stars. We detect the secondary star in every CV, and find evidence
for carbon deficits in
nearly all of them. In addition, we detect \raisebox{.6ex}{13}CO for the first
time in a CV secondary star. We find that
in the case of MU Cen, the strength of the \raisebox{.6ex}{13}CO feature
suggests that the CNO cycle has run to
completion (\raisebox{.6ex}{12}CO/\raisebox{.6ex}{13}CO = 3.2). We also find a
wide range in the strengths of lines from such elements
as sodium, calcium, magnesium, silicon, and iron when compared to main
sequence stars of the {\it most
appropriate} spectral type. The origin of such a wide range of behavior is not
easily identifiable. The
detection of enhanced \raisebox{.6ex}{13}CO certainly suggests that CNO
processed material has made it into the
atmospheres of a number of CV secondary stars, but whether this is from the
accretion of material,
or due to the evolutionary history of the secondary star itself, remains
unclear. Further high resolution,
high S/N spectroscopic observations will be needed to quantify these
anomalies. Equally important,
however, will be the need for good atmosphere models to help rule out any
effects due to peculiar
excitation conditions.

In the next section we discuss our observations, followed by a
description of the spectra of the
objects in section 3, followed by our conclusions in section 4.

\section{Observations}

Infrared spectroscopy for the program objects was obtained using
SPEX\footnote{For more on SPEX go to http://irtfweb.ifa.hawaii.edu/Facility/spex/} on the Infrared
Telescope Facility on Mauna Kea, and at the Cerro Tololo Interamerican
Observatory using OSIRIS\footnote{For information on OSIRIS go to http://www.ctio.noao.edu/instruments/ir\_instruments /osiris/index.html}
on the Blanco 4 m telescope. The observing run with OSIRIS occurred on 2002
March 20 and 21.
There were two different observing runs with SPEX: 2002 April 6 and 7, and
2003 May 16 to 19.
Both instruments were used in the mode that provided the highest possible
resolution in the {\it K}-band.
For SPEX, this consisted of using the spectrograph in single-order mode with a
0.3" slit, giving a
dispersion of 5.51 {\AA}/pix. The spectra produced in this mode covered the
entire {\it K}-band, from 1.96 to
2.50 $\mu$m. OSIRIS was also used in single order, long-slit mode, with a 0.5"
slit, the F/7 camera, and
the grating in 3\raisebox{.6ex}{rd} order. The resulting dispersion was 3.70
{\AA}/pix. We selected the grating angle to
cover the spectral region from 2.09 to 2.40 $\mu$m. For the CTIO run, the
conditions were photometric,
and the seeing was excellent (with an average FWHM near 0.5"). Unfortunately,
the conditions at the
IRTF were not photometric on any of the six nights, with the combined loss of
two nights due to fog
and clouds. The first of the observing runs at the IRTF also had poor seeing
(FWHM $\geq$ 1.2''). The
poor seeing at the IRTF was partly due to the loss of the dome air
conditioning system. During the
2003 May IRTF run, however, the seeing was excellent, eventually reaching FWHM
$\leq$ 0.4" in the pre-dawn hours.

The observing procedure was nearly identical for both observing
runs. The spectra obtained
with OSIRIS used a script that took five individual exposures along the slit
each separated from the
proceeding by 8 arcseconds. For just about all of the cataclysmic variables,
the exposure times were
four minutes in length. For the data obtained using SPEX, a similar observing
routine was employed,
but one in which data at six separate positions along the slit were obtained.
Typical exposure times
with SPEX were three minutes. For all of the program objects observed using
OSIRIS, observations
of nearby A-stars were obtained just before, or after, the observational
sequence for each target to
correct for telluric features. For the SPEX runs, however, we switched to the
use of early G-dwarfs
for the correction of telluric features as outlined by Maiolino et al. (1996).
Due to the longer
observational sequences for the fainter sources, observations of telluric
standards were often obtained
in the middle of a sequence if the change in airmass was significant ($\Delta$AM
$\geq$ 0.1). Finally,
observations of a number of bright, late-type stars were obtained to act as
spectral-type templates.
While the exposure times for these bright objects were quite short, the same
scripts as used for the
CV observations were employed.

To remove the sky background and dark current from each exposure
with OSRIS (SPEX), we
subtracted the median of the other four (five) exposures obtained in that
observing sequence. This
process resulted in five (six) background-subtracted exposures from which the
spectra were extracted
using the normal IRAF methods. The spectra were wavelength calibrated by
extraction of an arc
spectrum at the position (aperture) of each spectrum.

After wavelength calibration, groups of spectra for a CV were
combined to form a single
spectrum, and this was divided by the most appropriate telluric standard. If
several of these divided
spectra were created, as was the case for some of the fainter CVs, they were
medianed together to
form a final spectrum for the source. The OSIRIS spectra, having been divided
by an A star, were then
multiplied by a blackbody of the appropriate temperature and {\it K}-band flux
to create flux-calibrated
spectra. We caution against any interpretation of the profile of the H I Br-$\gamma$
line emission in the CV
spectra obtained with OSIRIS. Due to the fact that A dwarfs were used for
telluric correction of the
OSIRIS data, all of which have significant H I Br-$\gamma$ absorption, excess, false H 
I emission is produced
upon division. We decided not to attempt to construct H I-free A stars for the
final telluric division
due to the fact that we were uninterested in the H I emission.

The SPEX data were reduced in the same fashion, but instead of
using A-type dwarfs, we
observed early G-type dwarfs. The procedure for the use of G-type dwarfs to
correct for the telluric
features in near-infrared spectra has been described by Maiolino et al.
(1996). G-stars are useful for
telluric correction because they have very few strong absorption lines in the
near-infrared. But simple
division of a program object's spectrum by a that of an early G-type dwarf
does leave residual features
due to H I absorption and weak metal lines in the G dwarf spectrum. Maiolino
et al. have developed
an IRAF routine that modifies a high resolution infrared spectrum of the Sun
for the radial and
rotational velocities of the telluric standard. It then smooths this spectrum
to the resolution of the
spectrograph used to observe the telluric standard. We followed their
procedure to correct the data
obtained using SPEX. Because some of the same spectral features are expected
to be present in both
the G-star and CV secondary star spectra, we felt that a test of how well such
features are eliminated
by this procedure was warranted. To perform this exercise, we observed the
proto-planetary nebula
CW Leo (IRC +10 216), an object whose infrared spectrum is almost completely
free of atomic or
molecular absorption features (though weak absorption features of CO and other
molecules can be
seen longwards of 2.29 $\mu$m). The final {\it K}-band spectrum of CW Leo is
presented in Figure 1. We have
highlighted the region around H I Br-$\gamma$ in the insert to show how successfully
the telluric absorption
is removed, as are features introduced by the division of the G-dwarf spectrum
(such as an H I Br-$\gamma$
``emission" line). In the final, high S/N spectrum, there are no residual
features more than a few
percent above or below the continuum.

The final, fully reduced (but unsmoothed) spectra obtained using
SPEX (SY Cnc, RU Peg,
CH UMa, TT Crt, AC Cnc, EM Cyg, V426 Oph, SS Cyg, and AH Her) are shown in
Figures 2 and
3. The spectra obtained using OSIRIS (for V442 Cen, MU Cen, TT Crt, and BV
Pup) are shown in
Figure 4. Note that there remains some low-level fringing in the OSIRIS
spectra, confined to the
region between 2.21 and 2.26 $\mu$m, that we were unable to remove. We present the
un-smoothed
spectra in these figures so that the reader can determine for themselves the
noise level, and the
strength and/or reality of various features as ascertained from the smoothed
spectra of the individual
CVs to be discussed in the next section. A journal of our observations is
presented in Table 1.

\subsection{Accounting for Orbital Smearing}

Due to the faintness of the program CVs, and the long observing
sequences necessary to
obtain useful data, the production of a final spectrum can be compromised by
the orbital motion of
the secondary: By simply medianing spectra from all of the data obtained for a
particular object, the
narrow features of the secondary star are smeared-out. To properly account for
this requires an
ephemeris and radial velocity curve for the CV, and then a Doppler correction
of each individual
spectrum before the final result can be produced from their median. We have
listed the orbital phase
coverage for the program objects in Table 1. For those objects with accurate
ephemerides, we list the
run in orbital phase covered by our observations. For V442 Cen such
ephemerides do not exist, and
thus only the percentage of an orbital period covered by our observations is
listed. Given an accurate
ephemeris, it is rather easy to correct the spectra for the orbital motion,
and this has been performed
(when necessary) for each of the objects except V442 Cen.

\section{The {\it K}-Band Spectra of CV Secondary Stars}

Before we enter into a discussion of the individual spectra, we
take a moment to address two
issues that frequently arise that attempt to cast doubts on efforts to
interpret the secondary star spectra
observed for many CV systems. The first issue is a claim that what appears to
be a relatively normal
stellar photospheric spectrum has, in fact, been generated by the accretion
disk in the CV system. The
second issue is that the contamination of the secondary star spectrum by the
light from accretion
processes is so great that it fills-in, or veils the various absorption lines
to the point where very little
useful information about the true nature of the secondary star can be gleaned.
For both issues, SS Cyg
is probably the most useful example to examine in that it has both a luminous
secondary star, a
luminous accretion disk, and has been intensively studied. The first issue is
the easier of the two to
deal with, and can be logically dismissed by examining the profiles of the
detected absorption lines
with respect to their apparent velocity broadening. High resolution optical
spectra of SS Cyg by
Mart{\`\i}nez-Pais et al. (1994) reveal an early K-type spectrum that appears
to supply 55\% of the R-band
luminosity. The spectral energy distribution de-convolution by Harrison et al.
(2000) found that a
normal main sequence K2V star at the distance of SS Cyg would supply about
60\% of the observed
R-band flux. For SS Cyg, analysis of the absorption line spectrum for
rotational broadening gives
values of V\raisebox{-.6ex}{rot}sin{\it i} = 87 $\pm$ 4 km s\raisebox{.6ex}{-1}
(Mart{\'\i}nez-Pais et al 1994), 99 $\pm$ 8 km s\raisebox{.6ex}{-1} (Echevarria et
al. 1989),
and 90 $\pm$ 10 km s\raisebox{.6ex}{-1 }(Friend et al. 1990). These values are
consistent with the predicted secondary star
rotation velocity using the observed (and estimated parameters) for SS Cyg: 85
$\pm$ 8 km s\raisebox{.6ex}{-1} (Echevarria
et al. 1989). The average of the measured semi-amplitudes of the radial
velocity curve for the
secondary star in SS Cyg is K\raisebox{-.6ex}{abs} = 154 $\pm$ 2 km
s\raisebox{.6ex}{-1}. 

Obviously, if a late-type stellar spectrum was to be emitted from
a point interior to the orbit
of the secondary star, its velocity would have to be higher than that observed
for the secondary star.
It is also quite easy to surmise that if the entire accretion disk was engaged
in producing the K-dwarf
spectrum, the resulting range in orbital velocities of the disk material would
have dramatic
consequences on the observed spectrum: the absorption lines should have {\it
similar} {\it profiles} to the
emission lines. For SS Cyg, the H profile is double-peaked with a separation
of 526 $\pm$ 14 km s\raisebox{.6ex}{-1
}between the peaks. Since the observed absorption lines are not doubled, one
must propose that the
false stellar absorption spectrum is generated by a one-sided, narrow annulus
in the accretion disk that
somehow manages to both mimic a late-type stellar spectrum, and produce more
luminosity than the
secondary star. If we choose this annulus to be on the outer edge of the disk
(to keep its velocity
broadening to a minimum), and let the disk extend to a radius that is 90\% of
the Roche L\raisebox{-.6ex}{1} point in
SS Cyg (50\% of its semi-major axis), using equations from Warner (1995a), we
derive a Keplerian
orbital velocity of {\it v}\raisebox{-.6ex}{orb} = 344 km s\raisebox{.6ex}{-1}
for the material in this annulus (using {\it i} = 40\raisebox{.6ex}{o}, from
Mart{\'\i}nez-Paiz
et al. 1994), more than twice the observed value of K\raisebox{-.6ex}{2}.

There is little doubt that the spectrum of the secondary star for
some CVs is heavily
contaminated by emission from other sources in the binary system. This is
mostly evident through the
deviation of the slope of the continuum from that of an uncontaminated
late-type star. Some of the
infrared spectra that we will discuss below appear to have continua that
differ from that of an isolated
late-type star. Since no models exist that have fully explained their
quiescent spectral energy
distributions, it is difficult to deconvolve the various contaminating
components in a CV system to
allow us to fully extract the underlying secondary star spectrum. Our use of
the highest available
resolution has helped diminish the contamination of the narrow photospheric
lines by the accretion
disk (or white dwarf), as can be seen by comparing the R $\approx$ 2,000 spectrum of SS
Cyg shown in
Figure 5, with the R $\approx$ 1,200 spectrum shown in Figure 3 of Dhillon \& Marsh
(1995).

The real issue is how the contamination limits our ability to
extract spectral types, or other
information from infrared spectra. For heavily contaminated sources, the
contamination dramatically
affects the spectrum, and must be accounted for before we can determine the
nature of the secondary
star. For the long period CVs presented in this paper, however, we find that
the deviation from a pure,
late-type stellar spectrum is generally quite minimal. Thus, it is rather
simple to use the relative
strengths of various absorption features to estimate the spectral type of the
secondary star. Note that
using the relative strengths of features located in close proximity to each
other greatly reduces the
effect of contamination because the relative amount of contamination due to
other sources does not
change significantly over a small wavelength interval. For example, there are
two spectral forms that
are expected to be the main components of any contaminating sources in CVs:
hot blackbody
emission (from the white dwarf and/or hot spot), and some type of power-law
source (from the
accretion disk). Combinations of both of these, or even several different such
components, are
probably the most likely contaminating sources. But to strongly affect two
closely spaced absorption
lines requires a contaminating continuum source that changes dramatically over
a short wavelength
range. Of the various possible contaminants, a pure blackbody will impart the
greatest change due to
fact that its F$_{\lambda} \propto ^{-4}$ dependence is much
steeper than either the expected steady-state accretion disk
spectrum where F$_{\lambda} \propto ^{-7/3}$, or the spectrum of
free-free emission (F$_{\lambda} \propto ^{-2}$). For a 20,000 K
blackbody
in the {\it K}-band, twice as much flux is emitted at 2.00 $\mu$m vs. 2.40 $\mu$m.  But when comparing two
nearby features, such as the Na I doublet (centered near 2.207 $\mu$m) to the Ca I
triplet (centered near
2.263 $\mu$m), the difference is only 3\%! Thus, even in situations where the
overall contamination level
is high, the {\it relative} contamination between closely spaced absorption
features is negligible. As long
as limited spectral ranges are used when comparing spectral features, their
relative strengths will be
valid indicators of the true nature of the underlying spectrum. This is the
process we will use below
to estimate spectral types, as well as to make statements about apparent
abundance
enhancements/deficits.

In what follows, we describe the {\it K}-band spectra for each
individual object. In most cases, the
spectra presented in Figures 2, 3, and 4 have been smoothed to improve their
S/N. We will compare
the smoothed CV spectra to those of identically smoothed spectral templates
that we have obtained
using OSIRIS or SPEX, or to those from the catalogue of {\it K}-band spectra
for normal stars by Wallace
\& Hinkle (1997). In our spectral analysis, we have used {\it K}-band line
identifications from Wallace,
Hinkle, \& Bernath (1996), and from Hinkle, Wallace, \& Livingston (1995). For
each of the CVs, we
have rotationally broadened the spectral type templates to match the observed,
or predicted, rotational
broadening of the secondary star. We order the following discussion by orbital
period from the longest
to shortest.

\subsection{V442 Centauri}

In the Ritter-Kolb (1998) catalogue, the orbital period of V442
Cen is listed as P\raisebox{-.6ex}{orb} = 11.0 hr.
In the Downes \& Shara (2001) {\it A Catalogue and Atlas of Cataclysmic
Variables,} however, there is
some doubt cast on the reliability of this period. Warner (1995a) shows that
systems with such long
periods probably have to have evolved secondaries in order to fill their Roche
lobes. The {\it K}-band
spectrum of V442 Cen obtained with OSIRIS, shown in Figure 6, is the median of
15 four minute
exposures. At first glance, the spectrum is unusual compared to the other
objects in our sample in that
it lacks strong absorption features, and redward of 2.3 $\mu$m the spectrum is
choppy. While the Na I
doublet is strong, suggesting a late type star, there appears to be little
evidence for \raisebox{.6ex}{12}CO absorption.
The key to deriving a spectral type is the hump at 2.317 $\mu$m. This feature is an
opacity minimum
that shows up in mid- to late-type G dwarfs. It arises because in the earlier
G-type dwarfs, a strong
pair of iron lines are present that absorb the blue half of this hump. In
later G dwarfs, the \raisebox{.6ex}{12}CO\raisebox{-.6ex}{(3,1)}
transition comes into play, and whittles away at the hump from the red side.
This is shown in Fig. 7,
where we have compared the red end of the {\it K}-band spectrum of V442 Cen to
a G3V and a G8V. The
hump from the opacity minimum appears to be stronger in V442 Cen than for any
normal main-sequence dwarf, suggesting weaker iron and/or
\raisebox{.6ex}{12}CO absorption than normal. Note that this feature
does not show up in G-giants/subgiants due to their much stronger
\raisebox{.6ex}{12}CO absorption. There is another
opacity minimum at 2.372 $\mu$m, caused by a gap in the CO absorption, that appears
to be even stronger in
V442 Cen. However, the S/N ratio in this region is quite low, and is probably
partly responsible for
its large observed amplitude. We assign a spectral type of G6 $\pm$ 2 for the
secondary star. After typing
the secondary star, examination of the spectrum reveals that the expected
atomic absorption features
are present at their normal levels with, perhaps, slight enhancements in Na
and Mg. As shown in Fig.
6, the continuum of V442 Cen is substantially redder than a G8V, and is even
redder than a K0V.
Given these peculiar features, additional high S/N observations of V442 Cen
are clearly warranted.

\subsection{SY Cnc}

SY Cnc has an orbital period of 9.120 hrs, and has been classified
as a member of the Z Cam
family of cataclysmic variables. Vande Putte et al. (2003) have used the
technique of skew-mapping
to extract an updated ephemerides for the system, a radial velocity curve,
derived a mass ratio of {\it q}
= 0.68, and a primary mass of M\raisebox{-.6ex}{1} = 1.54 $\pm$ 0.40
M\raisebox{-.6ex}{}. These values suggest a secondary star with a
mass near 1.0 M\raisebox{-.6ex}{}. From these parameters, we estimate a very
low rotational broadening of V\raisebox{-.6ex}{rot}sin{\it i}
= 45 km/s. Our data set consists of eighteen three minute exposures with SPEX,
and the final
smoothed spectrum is shown in Fig. 8. The AAVSO light curve data base
indicates that SY Cnc was
in outburst, with m\raisebox{-.6ex}{v} $\approx$ 11.5, at the time of our observations. 

The spectrum of SY Cnc is consistent with that of an early G-type
star. Careful comparison
of SY Cnc to the G1.5V (HR7503) and G3V (HR7504) templates indicates a
slightly better match
to the G1.5V. This would be consistent with the secondary star's mass noted
above. In addition to the
He I line at 2.058 $\mu$m, the He I triplet at 2.112 $\mu$m is also in emission. Except
for Si I, which appears
to be somewhat underabundant, all of the absorption features appear to be at
the proper strength for
an early G-type dwarf. The peak due to the opacity minimum at 2.317 $\mu$m is
stronger in SY Cnc than
either of the templates, suggesting somewhat weaker Fe I absorption, though
none of the other
prominent iron lines seem to be significantly weaker than those of the
templates. The final feature of
interest is the slope of the continuum: In its outburst state, the slope of
the {\it K}-band continuum in SY
Cnc is redder than the G-star templates, implying that the outburst accretion
disk has a significantly
flatter spectral slope than a 6,000 K blackbody.

\subsection{RU Peg}

RU Peg has an orbital period of 8.990 hrs and, with
K\raisebox{-.6ex}{2MASS} = 10.46, is one of the brightest
CVs in the infrared. The unsmoothed SPEX spectrum of RU Peg, shown in Fig. 9,
is the median of
twelve two minute exposures, and covers only 0.05 in orbital phase. Friend et
al. (1990) derived the
following systemic parameters: K\raisebox{-.6ex}{2} = 121 km
s\raisebox{.6ex}{-1}, V\raisebox{-.6ex}{rot}sin{\it i} = 80 km
s\raisebox{.6ex}{-1}, M\raisebox{-.6ex}{2} = 1.07 $\pm$ 0.02, and a
spectral type of K3. Friend et al. note that the spectral type and mass are
not consistent, and suggest
that secondary star in RU Peg probably has begun to evolve off of the main
sequence (see \S4.2).
Comparison of RU Peg to the spectra of early K-dwarfs indicates a best-fit
spectral type of K2.
Analysis of the strengths of the absorption lines indicates only two
anomalies: weak Si I and CO. The
strongest absorption features from both Si and CO appear to be about one half
the strength they
should be for a spectral type of K2. These deficits appear to be confirmed
from the preliminary
analysis of a {\it FUSE} spectrum of RU Peg by Sion et al. (2002), which
indicates that the white dwarf
in RU Peg is underabundant in both carbon (0.1 $\times$ solar) and silicon
(0.1 $\times$ solar).

\subsection{CH UMa}

As in the case of RU Peg, Friend et al. (1990) find a
cooler-than-expected spectral type (M0)
for the secondary star in CH UMa. For an orbital period of 8.232 hr, and
assuming a main-sequence
mass-radius relationship, a G-type star would have been predicted. An analysis
of archival IUE
spectra of CH UMa by Dulude \& Sion (2002) reveals a large N V/C IV ratio,
similar to those
described by G\"ansicke et al. (2003), suggesting an enhanced level of
nitrogen, and a deficit of carbon.
Our spectrum of CH UMa, which consists of twelve three minute exposures, is
presented in Fig. 10.
Friend et al. estimate that V\raisebox{-.6ex}{rot}sin{\it i} 45 km
s\raisebox{.6ex}{-1}, and we have chosen to broaden our template spectra
by 40 km s\raisebox{.6ex}{-1}. The spectral type of the secondary star and the
{\it K}-band continuum are consistent with
a K7 $\pm$ 2, except for the extremely weak CO features. A number of other
features may be unusual
(such Ti I at 2.28 $\mu$m, and \raisebox{.6ex}{13}CO\raisebox{-.6ex}{(2,0)}) but
await spectra with higher S/N. The relatively narrow, single-peaked H I and He
I emission lines are consistent with the low orbital inclination angle of {\it
i} = 21\raisebox{.6ex}{o} $\pm$
4 found by Friend et al.

\subsection{MU Centauri}

MU Cen is a well known CV having a radial velocity curve for the
secondary (Friend et al.
1990) from which an orbital period (8.208 hr), mass ratio ({\it q} = 0.83 $\pm$ 
0.15), and secondary star
rotational velocity (V\raisebox{-.6ex}{rot}sin{\it i} = 110 $\pm$ 15 km/s) were
derived. We have used the observed value for
V\raisebox{-.6ex}{rot}sin{\it i} to broaden the template spectra for
comparison with the spectrum of MU Cen presented in
Figure 11. This spectrum was constructed from data obtained with OSIRIS, and
is the median of 10
four minute exposures. Ten percent of an orbital period was covered by our
observations. The
continuum slope of MU Cen is identical to mid-K type stars suggesting little,
if any contamination.
The Na I doublet and the Ca I triplet have peculiar profiles, but have roughly
the correct depth for a
spectral type near K5. We assign a spectral type of K4 $\pm$ 1. This spectral type
is consistent with the
results of Friend et al. which indicated a spectral type earlier than K7V.
However, the \raisebox{.6ex}{12}CO features,
and the Mg I line (at 2.2814 $\mu$m) are too weak for a mid-K type star. In
addition, both silicon and iron
seem to be slightly enhanced.  To the blue side of the Na I doublet, at 2.201
$\mu$m, is an absorption
feature that appears in many of our CV spectra (it is seen in the spectrum for
V442 Cen, for example),
but one whose origin is not obvious. There is a Ti I line near this position
(2.2010 $\mu$m), but this would
require a much greater enhancement than is indicated by the other Ti I lines.
The most striking feature
in the spectrum of MU Cen is the strong
\raisebox{.6ex}{13}CO\raisebox{-.6ex}{(2,0)} bandhead at 2.345 $\mu$m. A check on
the reality of this
feature can be made by putting it in context with the sequence of features
running from \raisebox{.6ex}{12}CO\raisebox{-.6ex}{(3,1)}
(2.321 $\mu$m), the red Na I doublet (2.336 $\mu$m), and
\raisebox{.6ex}{12}CO\raisebox{-.6ex}{(4,2)} (2.354 $\mu$m). Given the local S/N in
this
region it is obvious that the strength of this feature could differ by $\pm$ 20\%,
but this would still indicate
an extreme enhancement of \raisebox{.6ex}{13}CO. Comparison of the MU Cen
spectrum to those of red giants with
strong \raisebox{.6ex}{13}C enhancements indicates a
\raisebox{.6ex}{12}C/\raisebox{.6ex}{13}C ratio close to that of CNO
completion.

\subsection{TT Crateris}

TT Crt is a long period (7.3 hr), high inclination ({\it i}
$>$ 50\raisebox{.6ex}{o}) dwarf nova studied by Szkody et al.
(1992). We have used their value of {\it q} = 0.8, to estimate a rotational
velocity for the secondary star
of V\raisebox{-.6ex}{rot}sin{\it i} = 109 km/s, and have applied this amount
of broadening to the template star spectra. TT
Crt was observed using both OSIRIS and SPEX. A single spectrum was produced
from the OSIRIS
data that is the median of 15 four minute exposures, and this was combined
with the SPEX spectrum
of TT Crt that is the median of twenty four 3 minute exposures. TT Crt is
faint (K\raisebox{-.6ex}{2MASS} = 13.19), and
the smoothed spectrum presented in Fig. 12 is the mean of the OSIRIS and SPEX
data sets.
Comparison of the Na I, Ca I, and Mg I (@ 2.281 $\mu$m) absorption features gives a
best-fitting spectral
type of K5V, consistent with the results of Skzody et al. (1992), though the
Ca I triplet is somewhat
too weak for this spectral type. There is almost no evidence for
\raisebox{.6ex}{12}CO absorption! As in the spectra for
V442 Cen and MU Cen, the absorption feature at 2.201 $\mu$m is quite strong. While
our final spectrum
of TT Crt is rather poor, absorption lines from Ti I, Si I, and Fe I appear to
be at their normal
strengths, while those of Al I might be slightly enhanced.

\subsection{AC Cnc}

AC Cnc is a rarely observed eclipsing, nova-like variable that has
an orbital period of 7.211
hr, and K\raisebox{-.6ex}{2MASS} = 12.59. Optical observations indicate a
late-G/early-K spectral type for the secondary
star, and a mass ratio of {\it q} = 1.24 $\pm$ 0.14 (Schlegel et al. 1984). The
spectrum presented in Fig. 13,
which consists of eighteen three minute exposures, is very poor due to the
presence of patchy clouds.
Assuming a normal level of CO absorption, the spectral type of the secondary
appears to be somewhat
later than K5. The continuum, however, appears to be redder than this, and the
spectral type could be
as late as M2 if the CO features are weaker than normal. Given its long
orbital period, a dynamically
unstable value for its mass ratio ({\it q} $>$ 1), and the suggestion of
a fairly late spectral type, AC Cnc
certainly warrants additional attention.

\subsection{EM Cyg}

EM Cyg is a bright (K\raisebox{-.6ex}{2MASS} =11.15), well-known Z
Cam-type dwarf nova with an orbital
period of 6.982 hr, and has a secondary star with a spectral type of K2-5.
Recently, North et al. (2000)
have found that the light from the system is contaminated by a third star,
possibly associated with EM
Cyg, that is nearly identical in brightness and temperature to the secondary
star! With this discovery,
North et al. were able to derive a revised value for the mass ratio of {\it q}
= 0.88 $\pm$ 0.05, down from the
dynamically unstable value of {\it q} = 1.26 found by Stover et al. (1981).
The {\it K}-band spectrum of EM
Cyg, is shown in Fig. 14, and consists of twelve three minute exposures. North
et al. (2000) found
a rotational broadening of 140 $\pm$ 6 km s\raisebox{.6ex}{-1 }for the secondary
star in EM Cyg, and we have broadened
the template spectra by this amount.

The spectrum of EM Cyg is actually surprisingly peculiar.
Comparison of all of the main
molecular and atomic absorption features indicates a spectral type that is
slightly earlier than K0. For
example, the Mg I and Al I lines near 2.11 $\mu$m, the sodium doublets, and the CO
features are slightly
weaker in strength than those of a K0V template. While the Ca I triplet in EM
Cyg is stronger than
that of a K0V. The continuum, however, is much redder than a K0V, and is
almost identical to that
of our K5V template. What seems most apparent is that all of the spectral
features seem to be broader
than those in the templates. We believe this is partly due to our radial
velocity correction for the
secondary star. Given that the third star in the system is not moving, its
spectral features would have
been blue-shifted by 155 km s\raisebox{.6ex}{-1} due to our correction for the
velocity of the actual secondary star.
We have tested this scenario assuming both stars have identical luminosities,
and find that the spectral
features are indeed broader, but not as broad as those observed here. If we
assume an early/mid-type
K star, then the lines of Al I and Ti I are weaker than they should be, even
given the excess
broadening of the absorption features. Another clearly detected anomaly is
\raisebox{.6ex}{13}CO, which is several
times stronger in the spectrum of EM Cyg than in that of the K3V template.

\subsection{V426 Oph}
V426 Oph is a bright (K\raisebox{-.6ex}{2MASS} = 10.33) Z Cam CV
with an orbital period of 6.848 hr. During
outburst, V426 Oph reaches m\raisebox{-.6ex}{V} = 10.9 (Warner 1995, p130).
During its standstills, it hovers near m\raisebox{-.6ex}{V}
= 11.9. Inspection of the AAVSO light curve database indicates V426 Oph had
m\raisebox{-.6ex}{V} $\approx$ 12.2 at the time
of our observations. Hessman (1988) has used time-resolved optical
spectroscopy of V426 Oph to
derive the orbital period, an updated ephemerides, a mass ratio ({\it q} =
0.78 $\pm$ 0.06), an orbital inclination
(59\raisebox{.6ex}{o} $\pm$ 6), a secondary star spectral type (K3), and primary
and secondary star masses (0.9 M\raisebox{-.6ex}{} and 0.7
M\raisebox{-.6ex}{}, respectively). We have used this solution to predict a
value for the secondary star rotational
velocity of V\raisebox{-.6ex}{rot}sin{\it i} = 116 km s\raisebox{.6ex}{-1}.
The unsmoothed spectrum presented in Fig. 15 is the median of six
two minute exposures. Due to the brief time interval spent observing V426 Oph
(0.03 in orbital phase,
$\Delta$K\raisebox{-.6ex}{2} = 9 km s\raisebox{.6ex}{-1}), no radial velocity
correction has been applied to its spectrum.

The slope of the continuum of V426 Oph is much flatter/redder than
those of early/mid-K
dwarfs, being similar to that of an M2V, except for the lack of a decline due
to water vapor absorption
at the red end of the {\it K}-band. The absorption features also show a
mixture of strengths that span these
two spectral classifications. For example, the Al I and Mg I lines near 2.11 $\mu$m
are quite weak,
suggesting an early M spectral type. The Na I doublet, Ca I triplet, and Mg I
(at 2.281 $\mu$m) features
have strengths similar to those of the K5V template. The CO features are much
too weak for any
spectral type later than K2V. Given our results for SY Cnc, and because V426
Oph was 1.2 mags
above minimum light, it seems likely that accretion disk contamination is
causing the redder-than-expected spectral slope in V426 Oph. This allows us to
reconcile our result with the K3 classification
of Hessman (1988). By assigning a spectral type of K5 to the secondary in V426
Oph, we can then
compare the relative strengths of various features. It is clear that the Al I
lines and CO features are
much weaker in the spectrum of V426 than they should if the secondary were a
normal K-dwarf, with
both about half as strong as normal.  All of the other atomic species (Si I,
Ti I, Mg I, and Fe I) seem
to be present at near-normal strengths.

\subsection{SS Cygni}
SS Cyg has an orbital period of 6.603 hr and, as discussed
earlier, has V\raisebox{-.6ex}{rot}sin{\it i} = 87 $\pm$ 4 km
s\raisebox{.6ex}{-1} (Mart{\'\i}nez-Pais et al. 1994). We have applied this
amount of rotational broadening to our template
spectra. Due to its brightness, the spectrum of SS Cyg (presented earlier in
Figure 5) is of very high
quality, and is composed of six three minute exposures. We have identified
many of the strongest
atomic and molecular lines for the K5V template shown in Fig. 5. Using the
ephemerides of
Mart{\'\i}nez-Pais et al. (1994), the mean orbital phase at the time of our
observations was 0.1, just past
inferior conjunction of the secondary star. At first blush, the spectrum of SS
Cyg appears to be that
of a early to mid K-type dwarf. But closer examination reveals several
peculiarities. For example, the
strengths of the Ca I triplet and the first overtone feature of
\raisebox{.6ex}{12}CO closely resemble those of a K2 V.
The Na I doublet, and the higher overtones of CO, however, are much closer in
strength to those of
a K5V. Throughout the spectrum, the lines of Mg I are very weak. Perhaps the
most unusual region
in the spectrum of SS Cyg is that between 2.08 and 2.13 $\mu$m. The three Al I
features (at 2.092, 2.110,
2.117 $\mu$m) are not at their proper strengths, with the 2.110 $\mu$m line being
invisible. Between the
2.092 Al I line, and the 2.106 Mg I line, there appears to be a doublet in the
spectrum SS Cyg. The
bluer member of this pair can be associated with an Fe I line, and its
strength is roughly consistent
with other nearby Fe I lines--but the strength of those lines would require
excess iron absorption. A
comparison of the strength of the Fe I features in the spectrum of SS Cyg
between 2.20 and 2.30 $\mu$m
shows that iron does not appear to be overabundant. A possible origin for the
red line of this doublet,
and a number of other weak features in this region, including some of the
overly strong Fe I lines, can
traced to the assignment of the isolated absorption feature at 2.127 $\mu$m to CN.
CN lines can then
match the red line of the unidentified doublet as well as previously
unidentified absorption features
at 2.077 and 2.085 $\mu$m. As shown by the templates, CN absorption is not expected
to be significant
in early/mid K-type dwarfs.

Two other features of the SS Cyg spectrum deserve mention. The
first is the presence of a
weak \raisebox{.6ex}{13}CO\raisebox{-.6ex}{(2,0)} feature. While much weaker
than that seen in MU Cen, the S/N of the SS Cyg spectrum
is much higher, and this feature is certainly real. The other interesting
aspect of the spectrum of SS
Cyg is the overall slope of the continuum. The continuum of SS Cyg is flatter
than those of mid-K
dwarfs and, like SY Cnc and V426 Oph, there must be a significant
flat-spectrum, or additional red
source, contaminating the K-band spectrum.

Given all of these peculiarities it is difficult to assign a
unique spectral type to the secondary
star of SS Cyg.  We propose a spectral type of K4 $\pm$ 2 for the secondary star
in SS Cyg. This spectral
type should be compared with previous estimates of K4 (Harrison et al. 2000),
K2V (Echevarria et
al. 1989), K2/3 (Mart{\'\i}nez-Pais et al. 1994), and K5V (Friend et al.
1990).

\subsection{BV Puppis}
BV Pup is a dwarf novae that has an orbital period of 6.353 hr,
and exhibits low amplitude
outbursts. Modeling the infrared ellipsoidal variations of BV Pup, Szkody \&
Feinswog (1988) derived
a high inclination of 78\raisebox{.6ex}{o} $\leq$ {\it i} $\leq$ 90\raisebox{.6ex}{o} for
the system, and derived a mass of 0.62 M$_{\sun}$ for the secondary
star. This contrasts with the analysis by Bianchini et al. (2001) who derive a
much lower inclination
of 23\raisebox{.6ex}{o} $\pm$ 3 from a combination of spectroscopic observations
and dynamical arguments.  Bianchini
et al. derive a much higher secondary star mass of 0.96 M$_{\sun}$,
and a mass ratio of {\it q} = 0.80. They could
not detect any evidence for features from the secondary star in their optical
spectrum of this system.
We used those results to estimate a rotational velocity broadening of 61 km
s\raisebox{.6ex}{-1} for the BV Pup
secondary.  We observed BV Pup using OSIRIS, and the final spectrum,
consisting of ten four minute
exposures, is presented in Fig. 16.

The spectrum for BV Pup has rather low S/N. The slope of the
continuum suggests an early
to mid-K dwarf. Comparison of the relative strengths of the Na I and Ca I
features indicates a spectral
type near K3, as do the \raisebox{.6ex}{12}CO features. A normal, main
sequence K3V should have a mass near 0.72 M$_{\sun}$, suggesting the higher inclination model for BV Pup might
be more appropriate. A higher
inclination angle implies a larger rotational velocity, and this also seems to
be consistent with the
widths of the absorption features in the spectrum of BV Pup which are broader
than those of the
modified template stars. The spectrum is too poor for an abundance analysis,
though several lines of
Mg I seem stronger than those of the K3V template.

\subsection{AH Her}

AH Her is a Z Cam system with an orbital period of 6.195 hr that
varies between a quiescence
state where m\raisebox{-.6ex}{v} = 14.3, a standstill level near
m\raisebox{-.6ex}{v} = 12.0, and outburst at m\raisebox{-.6ex}{v} = 11.3
(Warner 1995b).
The AAVSO light curve data base indicates that AH Her was at
m\raisebox{-.6ex}{v} $\approx$ 12.5 at the time of our SPEX
observations, and on its way to a short-lived outburst that peaked about two
days later at m\raisebox{-.6ex}{v} = 11.5.
Horne et al. (1986) have developed a dynamical model for the system, including
a determination of
the secondary star rotation velocity: V\raisebox{-.6ex}{rot}sin{\it i =} 112$\pm$ 
17 km s\raisebox{.6ex}{-1}. 

AH Her is relatively bright (K\raisebox{-.6ex}{2MASS} = 11.38),
and the unsmoothed spectrum presented in Fig.
17 is the median of twelve three minute exposures. The spectrum of AH Her is
perhaps the most
unusual of the present survey. The continuum is slightly flatter than those of
K dwarfs, suggesting a
spectral type near M0. Given our results for SY Cnc, V426 Oph and SS Cyg,
however, this flat
spectrum is probably due to accretion disk contamination since AH Her was on
its way to an outburst.
All of the absorption features in the spectrum of AH Her are weaker than they
should be for a late-type star, suggesting that the contamination from the
accretion disk has seriously diluted the {\it K}-band
spectrum. If we use nearby spectral lines, we find that the ratios of most
strong atomic lines are
consistent with a mid-K spectral type. This agrees with the result from Horne
et al. (1986). However,
there are a number of anomalies, such as the missing Mg I doublet at 2.106 $\mu$m,
and the Al I line at
2.116 $\mu$m. It is probable that there is He I emission (at 2.113 $\mu$m) that is
filling-in these lines, as the
Mg I line at 2.281 $\mu$m is at its proper strength (relative to the nearby Ca I
triplet). At 2.288 $\mu$m,
between the Mg I line, and the first overtone of CO, is an unidentified
absorption feature. The entire
set of \raisebox{.6ex}{12}CO features are very weak, at most one-half their
normal strength. At the same time,
however, the \raisebox{.6ex}{13}CO\raisebox{-.6ex}{(2,0)} seems to be
detected! While the S/N in this region is not especially high, the
Na I doublet that precedes the \raisebox{.6ex}{13}CO\raisebox{-.6ex}{(2,0)} is
at a strength that is consistent with the other atomic lines
in the spectrum, and the \raisebox{.6ex}{13}CO\raisebox{-.6ex}{(2,0)} is at
least as strong as this doublet. New {\it K}-band spectra when AH
Her is in quiescence would be extremely valuable for further examination of
these anomalies.

\section{Discussion}

There are two consistent results from our {\it K}-band survey of
long period cataclysmic variables.
The first is that just about all of the CV systems studied here exhibit
weaker-than-expected CO
absorption features for their apparent spectral types. The other common trend
is for systems with
luminous accretion disks to have redder continuua than expected. There are two
plausible
explanations for the weakness of the CO features: either carbon and/or oxygen
are deficient in the
secondary stars of these systems, or weak CO emission is occurring from
elsewhere in the system that
fills-in the absorption features of the secondary star.  Howell et al. (2004)
have recently used Keck
to obtain high S/N observations of WZ Sge which reveal both CO and molecular
hydrogen (at 2.22 $\mu$m) in emission. In the case of WZ Sge, the H\raisebox{-.6ex}{2} emission was
about one third the strength of the
emission from the first overtone bandhead of CO. There is no evidence for
H\raisebox{-.6ex}{2} emission in any of the
objects discussed here. Given this, and the evidence for carbon deficits seen
in the UV spectra of
some of the white dwarfs in CVs (e.g., U Gem), it seems more likely that the
carbon is underabundant
in CV secondary stars. Unfortunately, because the CVs in our sample are all
long period systems with
G and K-dwarf secondaries, the water vapor features seen in the spectra of the
M-type stars found in
short-period CVs are not present, and it will be difficult to conclusively
rule out an oxygen deficit for
these objects using {\it K}-band spectra.  

That the {\it K}-band continua of many long period CVs appear
flatter/redder than the spectra of
G and K-type secondaries can be explained by having a contaminating source
that has a spectrum that
is less steep than the F$_{\lambda} \propto \lambda^{-4}$, blackbody-like
spectra of G and K-type stars. Both the standard
accretion disk spectrum (F$_{\lambda} \propto \lambda^{-7/3}$) and
free-free emission (F$_{\lambda} \propto \lambda^{-2}$) are possible
sources. To
demonstrate what happens when we remove a contaminating source, we present the
spectrum of V426
Oph in Fig. 18 from which a flat continuum source (F$_{\lambda}$ =
constant), with 44\% of its {\it K}-band flux, has
been subtracted. While the resulting spectrum has the identical slope to the
K5V template, the
absorption lines of sodium and calcium are now too strong for a dwarf star! To
achieve the same
result with an free-free or accretion disk spectrum requires them to
constitute an even larger fraction
of the {\it K}-band flux, with the result of even stronger absorption lines.
Only the subtraction of spectra
with a positive spectral index can minimize this effect. We are unaware of any
physical process that
generate such spectra, but observations using SIRTF would help quantify the
nature of this emission.

\subsection{$^{13}$CO and Other Apparent Abundance Anomalies}
The goal of our program was to investigate whether the secondary
stars in long period CVs
show evidence for peculiar abundances. We summarize our results in Table 2,
where a ``+" indicates
a possible enhancement, and a ``-" a deficit (a ``?" indicates uncertain,
while an ``!" indicates a
significant deficit). Ellipsis indicate the spectrum was either too poor, or
too contaminated, to have
confidence in statements about the abundance of a particular element. A ``0"
indicates that a species
seems to be at relatively normal level. While every single object appears to
have something peculiar
about its spectrum, the {\it K}-band data alone are not quite sufficient to
determine the source of these
peculiarities. Only for SS Cyg, RU Peg, and V426 Oph, were the S/N of the
spectra sufficiently high
to confidently examine them for low-level enhancement/deficits. 

Except for the near-universal weakness of CO, there is no apparent
pattern in the strength of
the lines from any particular element.  This indicates to us that normal
cosmic dispersion might be
mostly responsible for the observed abundance peculiarities. Given the complex
environment in
which the spectrum of a CV is emitted, unusual line strengths could also arise
due to other effects.
One feature, however, does stand out: the detection of \raisebox{.6ex}{13}CO
in four of the systems (MU Cen, EM Cyg,
SS Cyg, and AH Her). In Fig. 19, we show close-up views of the spectra of
these four objects in the
region around the \raisebox{.6ex}{13}CO\raisebox{-.6ex}{(2,0)} bandhead. In 
the normal main-sequence counterparts of the (mostly) K-type stars
found in this sample of long-period CVs, any \raisebox{.6ex}{13}CO absorption
is almost undetectable. But we find
evidence for fairly strong features in these four CVs. Even in the objects
where the S/N of the spectra
are quite low (e.g., CH UMa), there is evidence for \raisebox{.6ex}{13}CO
features. {\it Only for RU Peg and V426 Oph
can we rule out enhanced levels of \raisebox{.6ex}{13}CO.} In the models of
Marks \& Sarna (1998), the presence of \raisebox{.6ex}{13}C
was strong evidence for evolution of the secondary star, resulting from the
baring of, or mixing-in
material from, layers where the CNO cycle had been operating. In such cases,
the isotopic ratios of
carbon, nitrogen and oxygen in the photospheres of the secondary star were
found to reach unusually
large values. In this process, the common isotopes
(\raisebox{.6ex}{12}C,\raisebox{.6ex}{ 14}N, and \raisebox{.6ex}{16}O) of
these three species become
depleted. Given that we simultaneously observe both an apparent
\raisebox{.6ex}{12}CO depletion {\it and} \raisebox{.6ex}{13}CO
enhancement suggests that material enriched from the CNO cycle is reaching the
photospheres of CV
secondary stars.

As described earlier, there are at least four presently envisioned
paths that can provide CNO
isotopic enrichment: 1) the accretion of CNO processed material during the
common envelope phase,
2) the accretion of novae ejecta, 3) the possibility that the donor star
originally had a higher mass than
the white dwarf primary, and that it is now the CNO-enriched core of this
massive star, and 4) the
secondary stars to have begun to evolve off of the main sequence before
becoming a CV. The first
two paths are consistent with the current evolutionary paradigm for CVs that
requires the secondary
stars have undergone little evolution during their lifetimes as short period
binaries. The scenario
where the secondary stars suffers large CNO enrichment from the accretion of
novae ejecta seems
difficult to sustain, since the time between classical novae explosions is
large ($\sim$ 10\raisebox{.6ex}{4} yr) and the amount
of material that can be accreted is relatively small ($\ll$
10\raisebox{.6ex}{-4} M$_{\sun}$). The small amount of ejecta that
could be realistically accreted would get mixed-in to the secondary star
quickly enough to be become
virtually undetectable. A similar scenario is envisaged for the common
envelope phase. The time
interval between the common envelope phase and the time of first contact of
the secondary star with
its Roche lobe is believed to be so long ($>$ 10\raisebox{.6ex}{8} yr),
that any accreted material would get thoroughly
mixed-into deeper layers within the secondary star.

The apparent detection of enhanced levels of \raisebox{.6ex}{13}CO
indicates that CNO processed material is
present in the atmospheres secondary stars of these long period systems, and
that they have either
undergone some evolution off of the main sequence, or they are the stripped
cores of more massive
stars.  The main difficulty with either of these two scenarios is that
population synthesis models by
Howell et al. (2001) find that very few CVs are formed with high mass
secondary stars. It is
interesting to note that we {\it do} detect at least one relatively normal CV
secondary star (SY Cnc) that
has a spectral type similar to the sun. The possibility there are some CV
secondary stars with initial
masses of $>$ 1 M$_{\sun}$, does not seem too far-fetched.

\subsection{The Case for Subgiant Secondary Stars in SS Cyg and RU Peg}

In Harrison et al. (2000), a investigation into the nature of the
secondary star in SS Cyg was
made by combining a high-precision {\it HST Fine Guidance Sensor} (FGS)
parallax with ground-based
photometry. They found that if all of the infrared luminosity of SS Cyg was
presumed to be coming
from the secondary star, then the secondary star must be a K4 subgiant.
Recently, Harrison et al.
(2004) have reported an {\it HST} FGS parallax for RU Peg ($\pi$ = 3.55 mas). Given
that the spectrum of
the secondary star in RU Peg appears uncontaminated by accretion disk
emission, we thought it
interesting to compare it to SS Cyg. In Fig. 20, we present the SEDs of SS Cyg
(data from Harrison
et al. 2000) and RU Peg (data from Harrison et al. 2004). The SEDs of the two
systems are very
similar. Their observed K magnitudes differ by $\Delta$K = 1.12 mag, while their
distance moduli differ by
1.16 mag! Thus, these two systems have nearly identical {\it K}-band
luminosities: M\raisebox{-.6ex}{K} = 3.26 (RU Peg),
and M\raisebox{-.6ex}{K }= 3.30 (SS Cyg). These values should be compared to a
K2V (the derived spectral type for
RU Peg), which has an absolute magnitude of M\raisebox{-.6ex}{K }= 4.15, while
a K4V (the spectral type for SS Cyg)
has M\raisebox{-.6ex}{K} = 4.48. Thus, if the entire {\it K}-band luminosities
are ascribed to their secondary stars, RU Peg
is 0.89 mags above the main sequence, and SS Cyg is 1.18 mags above the main
sequence. It is
difficult to envision a scenario where accretion disk contamination could
supply $\approx$ 1 mag of
luminosity, yet not severely contaminate the secondary star spectrum. It
therefore seems quite likely
that both RU Peg and SS Cyg have secondary stars that have evolved off of the
main sequence.

\section{Conclusions}
We have obtained moderate resolution {\it K}-band spectra of a
dozen long period (P\raisebox{-.6ex}{orb} $>$ 6 hr)
cataclysmic variables and clearly detect the secondary stars in every system.
We find weaker than
normal \raisebox{.6ex}{12}CO absorption in nearly every object. There is
evidence for the enhancements/deficits for
other elements, but for the most part, the spectra lack sufficient S/N to make
conclusive statements.
Higher S/N data are clearly needed, but will require 8 and 10 m telescopes. In
addition, it would be
extremely useful to have somewhat higher resolution data to carefully examine
the CO features to
determine if there is low-level CO emission occurring that might create the
false appearance of weak
absorption features. Mid-infrared photometry, such as provided by SIRTF, would
help unravel the
redder-than-expected continuua, allowing us to deconvolve, and remove, the
contaminating flux from
the spectra of systems with luminous accretion disks. 

~

\noindent
Acknowledgments: This research was performed with funding from National
Science Foundation
grant AST-9986823, which provided partial support for TEH. HLO acknowledges
support from a
New Mexico Space Grant Consortium fellowship. This research has made use of
the NASA/IPAC
Infrared Science Archive, which is operated by the Jet Propulsion Laboratory,
California Institute of
Technology, under contract to te National Aeronautics and Space
Administration. This publication
also makes use of data products from the Two Micron All Sky Survey, which is a
joint product of the
University of Massachusetts and the Infrared Processing and Analysis
Center/California Institute of
Technology, funded by the National Aeronautics and Space Administration and
the National Science
Foundation. We would also like to acknowledge use of the AAVSO's online data
archive to examine
the outburst states of the program CVs.

\begin{flushleft}
{\bf References}

Bianchini, A., Skidmore, W., Bailey, J. M., Howell, S., \& Canterna, R. 2001,
A\&A, 367, 588

Cheng, F. H., Sion, E. M., Szkody, P., Huang, M. 1997, ApJ, 484, L149

Dhillon, V. S., \& Marsh, T. R. 1995, MNRAS, 275, 89

Downes, R. A., \& Shara. M. M. 2001, PASP 113, 764

Dulude, M., \& Sion, E. M. 2002, BAAS 34, 1163

Echevarria, J., Diego, F., Tapia, M., Costero, R., Ruiz, E., Salas, L.,
Gutierrez, L., Enriquez, R. 1989,
MNRAS, 240, 975

Friend, M. T., Martin, J. S., Connon-Smith, R., Jones, D. H. P. 1990, MNRAS,
246, 654

Haefner, R., Metz, K. 1982, A\&A 109, 171 

Haefner, R., Betzenbichler, W. 1991, IBVS No. 3665

Harrison, T. E., McNamara, B. J., Szkody, P., Gilliland, R. L. 2000, AJ 120,
2649

Harrison, T. E., Johnson, J. J., McArthur, B. E., Benedict, B. J., Szkody, P.,
Howell, S. B., \& Gelino,
D. M. 2004, AJ 127, 460

Hessman, F. V. 1988, A\&A Supp., 72, 515

Horne, K., Wade, R. A., \& Szkody, P. 1986, MNRAS, 219, 791

Howell, S. B., Harrison, T. E., Szkody, P. 2004, ApJ, 602, L49

Howell, S. B., Nelson, L. A., Rappaport, S. 2001, ApJ, 550, 897

Jos\'e, J., Coc, A., \& Hernanz, M. 2001, ApJ, 560, 897

Long, K., \& Gilliland, R. L. 1999, ApJ, 511, 916

Lyons, K., Stys, D., Slevinsky, R., Sion, E., Wood, J. H. 2001, AJ, 122, 327

Maiolino, R., Rieke, G. H., \& Rieke, M. J. 1996, AJ, 111, 537

Marks, P. B., \& Sarna, M. J. (1998) (MS98), MNRAS, 301, 699

Mart{\'\i}nez-Pais, I. G., Giovannelli, F., Rossi, C., Gaudenzi, S. 1994,
A\&A, 291, 455

Mennickent, R. E., \& Diaz, M. P. 2002, MNRAS, 336, 767

North, R. C., Marsh, T. R., Moran, C. K. J., Kolb, U., Smith, R. C., \&
Stehle, R. 2000, MNRAS, 313,
383

Ritter, H. \& Kolb, U. 1998, A\&AS, 129, 83

Schlegel, E. M., Kaitchuck, R. H., \& Honeycutt, R. K. 1984, ApJ, 280, 235 

Sion, E. M., Cheng, F., Godon, P., \& Szkody, P. 2002, BAAS 34, 1162

Sion, E. M. 1999, PASP, 111, 532

Sion, E. M., Cheng, F. H., Sparks, W. M., Szkody, P., Huang, M., Hubeny, I.
1997, ApJ, 480, L17 

Sion, E. M., Cheng, F. H., Szkody, P., Sparks, W., Gaensicke, B., Huang, M.,
Mattei, J., 1998, ApJ,
496, 449

Sion, E. M., Szkody, P., Gaensicke, B., Cheng, F. H., LaDous, C., Hassall, B.
2001, ApJ, 555, 834

Stover, R. J. 1981, ApJ, 249, 673

Szkody, P., Williams, R. E., Margon, B., Howell, S. B., \& Mateo, M. 1992,
ApJ, 387, 357

Szkody, P., \& Feinswog, L. 1988, ApJ, 334, 442

Urban, J.., Lyons, K., Mittal, R., Nadalin, I., DiTuro, P., Sion, E 2000,
PASP, 112, 1611

Vande Putte, D., Smith, R. C., Hawkins, N. A., \& Martin, J. S. 2003, MNRAS,
342, 151

Wallace, L., \& Hinkle, K. 1997, ApJS, 111, 445

Wallace, L., Hinkle, K. \& Bernath, P. 1996, ApJS, 106, 165

Hinkle, K., Wallace, L., \& Livingston, W. 1995, PASP, 107, 1042

Warner, B. 1995a, {\it Cataclysmic Variable Stars}, (Cambridge University
Press:Cambridge, England),
p109.

Warner, B. 1995b, {\it Cataclysmic Variable Stars}, (Cambridge University
Press:Cambridge, England),
p130.
\end{flushleft}

\begin{deluxetable}{cccccc}
\tablecaption{Observation Journal}
\tablehead{\colhead{Object} &\colhead{Instrument} & \colhead{Date Obs.}
&\colhead{Start (UT)}  &  \colhead{Stop (UT)} & \colhead{Phase}
}
\startdata
V442 Cen& OSIRIS& 2002 Feb. 21 &  5:12 &  6:28 & 11.5\%  \\
SY Cnc  &  SPEX & 2003 May 19  & 5:20  & 6.23  &  0.64 - 0.75\\
RU Peg  & SPEX  & 2003 May 18  & 14:39 & 15:08 &  0.63 - 0.68\\
CH UMa  & SPEX  & 2003 May 17  & 7:12  & 7:47  &  0.53 - 0.60\\
MU Cen  & OSIRIS& 2002 Feb. 22 & 7:51  & 8:40  &  0.03 - 0.13\\
TT Crt  & OSIRIS& 2002 Feb. 22 & 6:12  & 7:16  &  0.36 - 0.51\\
TT Crt  & SPEX  & 2003 May 17  & 8:08  & 9:34  &  0.24 - 0.43\\
AC Cnc  & SPEX  & 2003 May 18  & 5:22  & 6:54  &  0.64 - 0.75\\
EM Cyg  & SPEX  & 2003 May 18  & 12:21 & 13:05 &  0.08 - 0.19\\
V426 Oph& SPEX  & 2003 May 17  & 9:58  & 10:12 &  0.26 - 0.31\\
SS Cyg  & SPEX  & 2002 Apr. 7  & 15:05 & 15:46 &  0.74 - 0.84\\
BV Pup  & OSIRIS& 2002 Feb. 21 & 3:40  & 4:25  &  0.97 - 0.08\\
AH Her  & SPEX  & 2003 May 18  & 8:27  & 9:07  &  0.14 - 0.24\\
\enddata 
\end{deluxetable}

\begin{deluxetable}{ccccccccc}
\tablecaption{Apparent Abundance Anomalies}
\tablehead{\colhead{Object}  & \colhead{Na}  & \colhead{Mg}  & \colhead{Al}  & \colhead{Si}  &  \colhead{Ca}  & \colhead{Ti}  & \colhead{Fe}  &  \colhead{CO} 
} 
\startdata
V442 Cen & +? & +? & \nodata & \nodata  & \nodata  & \nodata  & $ -$? & $-$? \\ 
SY Cnc   &  0 &  0 &  0  & $-$  &  0  &  0  &  0  &  0  \\
RU Peg   &  0 &  0 & 0  & $-$  &  0  &  0  &  0  & $-$ \\
CH UMa   & \nodata & \nodata & \nodata & \nodata & \nodata & \nodata & \nodata & $-$! \\
MU Cen   & \nodata & \nodata & \nodata &  +  & \nodata & \nodata &  +  &  $-$, 13$^{+}$\\
TT Crt   & \nodata & \nodata &  +?   & \nodata & $-$?   & \nodata & \nodata & $-$!\\
AC Cnc   & \nodata & \nodata & \nodata & \nodata & \nodata& \nodata & \nodata &  0? \\
EM Cyg   & \nodata & \nodata & $-$ & \nodata & \nodata &$-$ & \nodata & $-$, 13$^{+}$ \\ 
V426 Oph   &  0  &  0  & $-$ & 0  &  0  &  0  &  0  & $-$ \\
SS Cyg   &  0  & $-$ & $-$?  &  0 &  0  &  0  &  0  & $-$, 13$^{+}$ \\ 
BV Pup   & \nodata &  +? & \nodata & \nodata & \nodata & \nodata & \nodata & 0
\\ 
AH Her   & \nodata & \nodata & $-$? & \nodata & \nodata & \nodata & \nodata &
$-$!, 13$^{+}$\\
\enddata
\end{deluxetable}

\clearpage
\begin{flushleft}
{\bf Figure Captions}

Figure 1. The {\it K}-band spectrum of proto-planetary nebula CW Leo (= IRC
+10 216). Shortward of
2.29 $\mu$m, the spectrum of CW Leo is free of strong absorption features. In the
inset, we show an
expanded view (the flux scale of the inset is twice that of the full spectrum)
of the region spanning
2.10 to 2.20 $\mu$m. In this inset, the bottom spectrum is the raw, extracted
spectrum before division
by the spectrum of a G-dwarf star. The middle spectrum is after division by
the G-dwarf, revealing
false emission lines due to weak absorption features in the G-star spectrum
(the strongest of these
is H I Br-$\gamma$ at 2.1655 $\mu$m). The final spectrum is constructed by the
multiplication of a modified
solar spectrum that has been smoothed to the resolution of the SPEX
instrument, and corrected for
the radial velocity of the template star using an IRAF routine developed by
Maiolino et al. (1996).

~

Figure 2. The final, but unsmoothed {\it K}-band spectra of SY Cnc, RU Peg, CH
UMa, and TT Crt
obtained using SPEX.

~

Figure 3. As in Figure 2, but for AC Cnc, EM Cyg, V426 Oph, SS Cyg, and AH
Her.

~

Figure 4. The final, unsmoothed OSIRIS spectra for V442 Cen, MU Cen, TT Crt,
and BV Pup.

~

Figure 5. The (unsmoothed) {\it K}-band spectrum of SS Cyg. We compare the SS
Cyg spectrum to that
of K2V and K5V templates from Wallace \& Hinkle (1997). We identify the most
prominent atomic
and molecular absorption features below the spectrum of the K5V template.

~

~

Figure 6. The spectrum of V442 Cen compared to those of a G8V and K0V. All
three spectra have
been smoothed to a FWHM resolution of 40 {\AA} using the ``gauss" routine in
IRAF.

~

Figure 7. A close-up of the red region of the V442 Cen spectrum (in green)
showing the location
of the two opacity minima described in the text. Also plotted are the {\it
K}-band spectra of G3V (blue)
and G8V (red) templates. In this figure, the individual lines from the
vibrational/rotational
transitions of the CO molecule are plotted as vertical dotted lines, Fe I
lines are indicated in red,
green lines are Na I, while blue lines are Ti I. In the hotter (G2V) star, the
Fe I lines absorb the
continuum at 2.317 $\mu$m, in later-type stars ($>$ G8V), the
\raisebox{.6ex}{12}CO\raisebox{-.6ex}{(3,1)} bandhead becomes stronger, and
completely removes this feature. The strength of the opacity minimum in V442
Cen is greater than
any normal G-dwarf, suggesting an iron and/or CO deficit, though neither of
these species appears
to be underabundant.

~

Figure 8. The spectrum of SY Cnc (smoothed to FWHM = 20 {\AA}) compared to the
spectra of a
G1.5V and a G3V template. SY Cnc also shows an apparent emission feature at
2.317 $\mu$m due to
the opacity minium described for V442 Cen in Fig. 7. The dotted vertical lines
in this plot denote
the locations of the strongest Si I lines in the {\it K}-band spectra of cool
stars. It is clear that the Si I
features in the spectrum of SY Cnc are weaker than those of the template
dwarfs.

~

Figure 9.  The unsmoothed spectrum of RU Peg compared to K2V and K3V
templates. As in Fig.
8, the dotted vertical lines are the locations of the Si I lines.

~

Figure 10. The smoothed (FWHM = 20 {\AA}) spectrum of CH UMa, compared to the
spectra of a
K5V template from Wallace \& Hinkle (1996), and that of an M2V (GJ393)
obtained using OSIRIS
(the region around H I Br-$\gamma$ at 2.1655 $\mu$m has been patched-over).

~

Figure 11. The spectrum of MU Cen compared to those of K3V and K5V templates.
The location
of the \raisebox{.6ex}{13}CO bandhead is marked. All three spectra have been
smoothed to FWHM = 20 {\AA}. The
vertical dotted line is the location of the 2.281 $\mu$m Mg I line, which appears
to be extremely weak
in the spectrum of MU Cen. The arrow points to the location of an unidentified
absorption feature
(at 2.201 $\mu$m) seen in the spectra of several of our program objects.

~

Figure 12. The spectrum of TT Crt. This spectrum is the mean of data sets from
both OSIRIS and
SPEX, and has been smoothed to FWHM = 40 {\AA}, and compared to identically
smoothed K5V and
M2V spectral type templates. As in Fig. 11., the location of an unidentified
line at 2.201 $\mu$m is
indicated with an arrow.  

~

Figure 13.  The spectrum of AC Cnc compared to K3V and K5V templates. The
three spectra have
been smoothed to FWHM = 40 {\AA}.

~

Figure 14.  The (unsmoothed) spectrum of EM Cyg compared to those of K0V and
K3V templates.
The vertical dotted lines indicate the locations of the strongest Al I lines
in the {\it K}-band spectra of
cool dwarfs. 

~

Figure 15. The unsmoothed spectrum of V426 Oph compared to K5V and M2V
templates. The
slope of the continuum in V426 Oph is the flattest of any of the twelve CVs
presented in this paper.
As in Fig. 14, the dotted vertical lines mark the locations of the strongest
Al I lines.

~

Figure 16. The smoothed (FWHM = 40 {\AA}) spectrum of BV Pup compared to
identically processed
K2V and K5V templates. Like the other {\it K}-band spectra obtained using
OSIRIS, the H I Br-$\gamma$
emission line at 2.1655 $\mu$m is artificially enhanced due to the division of an
A-type dwarf. The
vertical dotted lines in this plot are the locations of the strongest Mg I
lines in the spectra of late-type dwarfs. The strength of these lines seem to
be slightly enhanced in the spectrum of BV Pup.

~

Figure 17.  The unsmoothed spectrum of AH Her, compared to K3V and K5V
templates. The Mg
I doublet at 2.106 $\mu$m, and the Al I line at 2.116 $\mu$m are identified with arrows.
Both features are
very weak in the spectrum of AH Her. An unidentified absorption feature at
2.288 $\mu$m is also
marked. While the \raisebox{.6ex}{12}CO features for AH Her are very weak, the
\raisebox{.6ex}{13}CO\raisebox{-.6ex}{(2,0)} bandhead at 2.345 $\mu$m
(dotted line) is clearly present.

~

Figure 18. The spectrum of V426 Oph (bottom) is much flatter than that of a
K5V template
(middle), even though most of the absorption features are consistent with an
early/mid-type K
dwarf. After subtraction of a contaminating source that has a flat spectrum
(F$_{\lambda}$ = constant) and that
supplies 44\% of the {\it K}-band flux, the slope of the continuum of V426 Oph
(top) now matches that
of the K5V template. Now, however, all of the absorption features (except CO)
are now much
stronger than seen in a K5V.

~

Figure 19. A close-up view of the spectra of MU Cen, EM Cyg, SS Cyg and AH Her
showing the
region of the {\it K}-band containing the main CO features. The dotted
vertical lines indicate the location
of the bandheads for \raisebox{.6ex}{12}CO\raisebox{-.6ex}{(2,0)} at 2.294 $\mu$m,
\raisebox{.6ex}{12}CO\raisebox{-.6ex}{(3,1)} at 2.321 $\mu$m, the
\raisebox{.6ex}{13}CO\raisebox{-.6ex}{(2,0)} feature at 2.345 $\mu$m,
and \raisebox{.6ex}{12}CO\raisebox{-.6ex}{(4,2)} at 2.354 $\mu$m. The {\it K}-band
spectra of normal K2 and K5 dwarfs are plotted for
comparison.

~

Figure 20. The observed spectral energy distribution for SS Cyg (solid
circles) from Harrison et al.
(2000), and RU Peg (stars).

\end{flushleft}
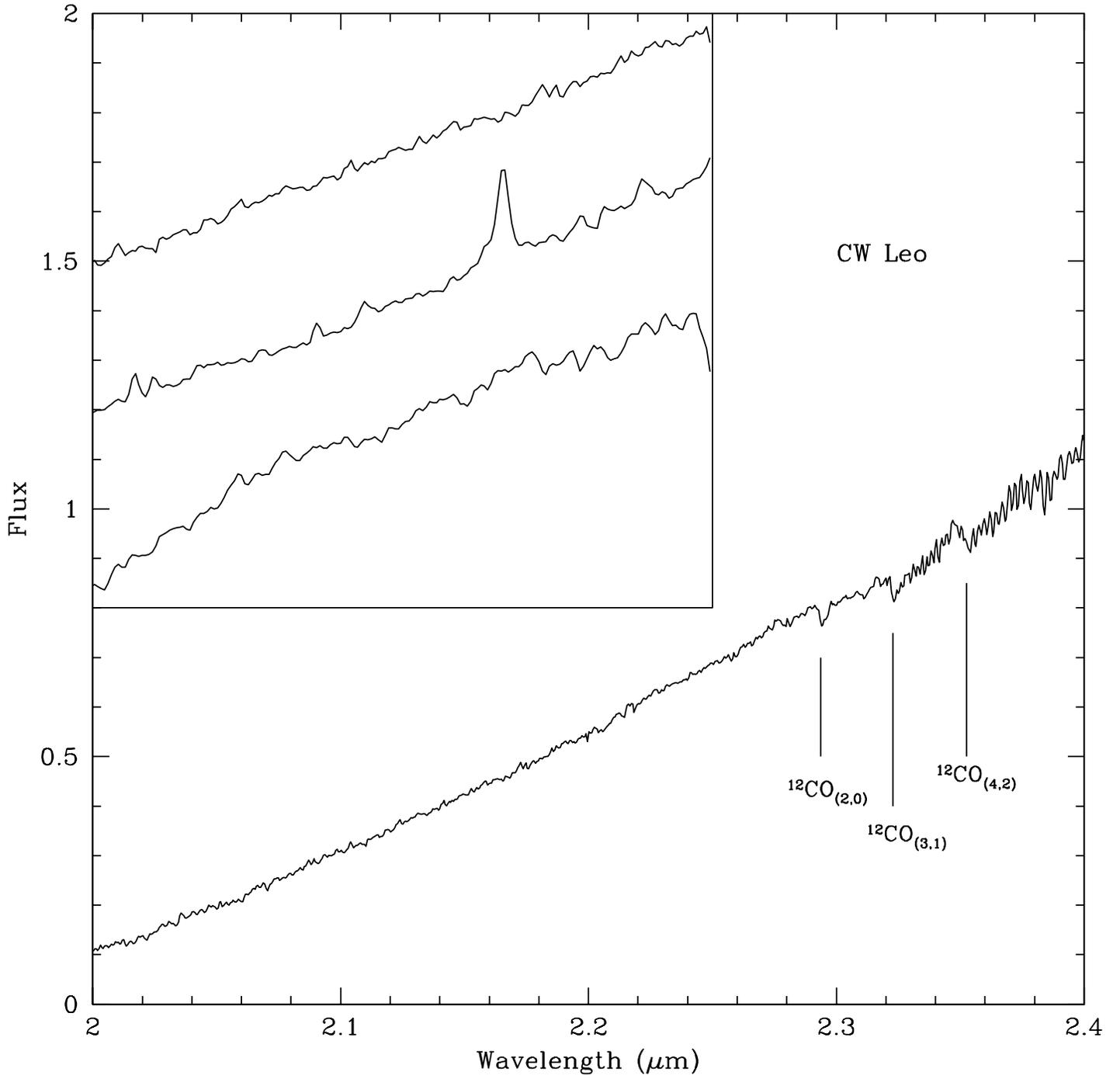
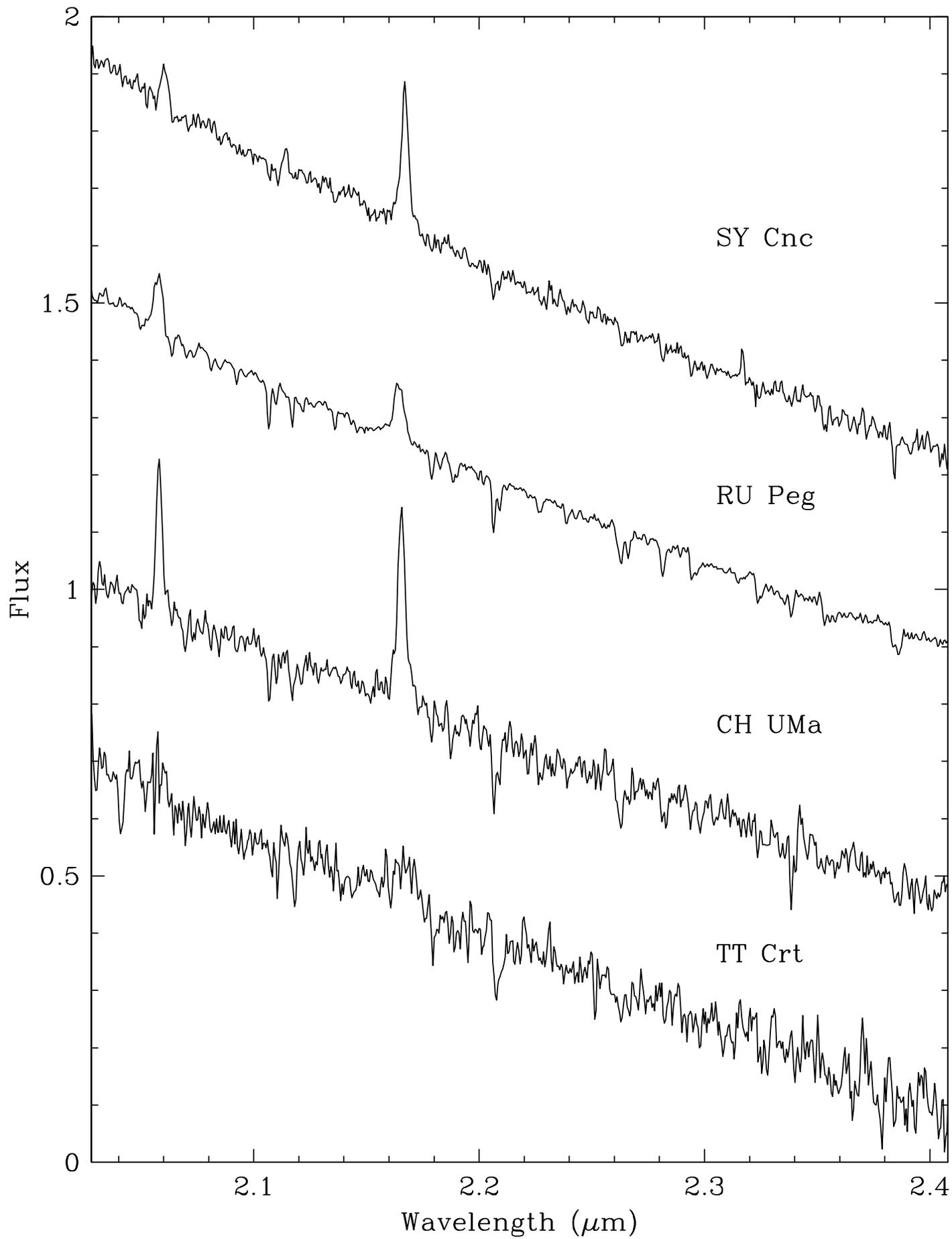
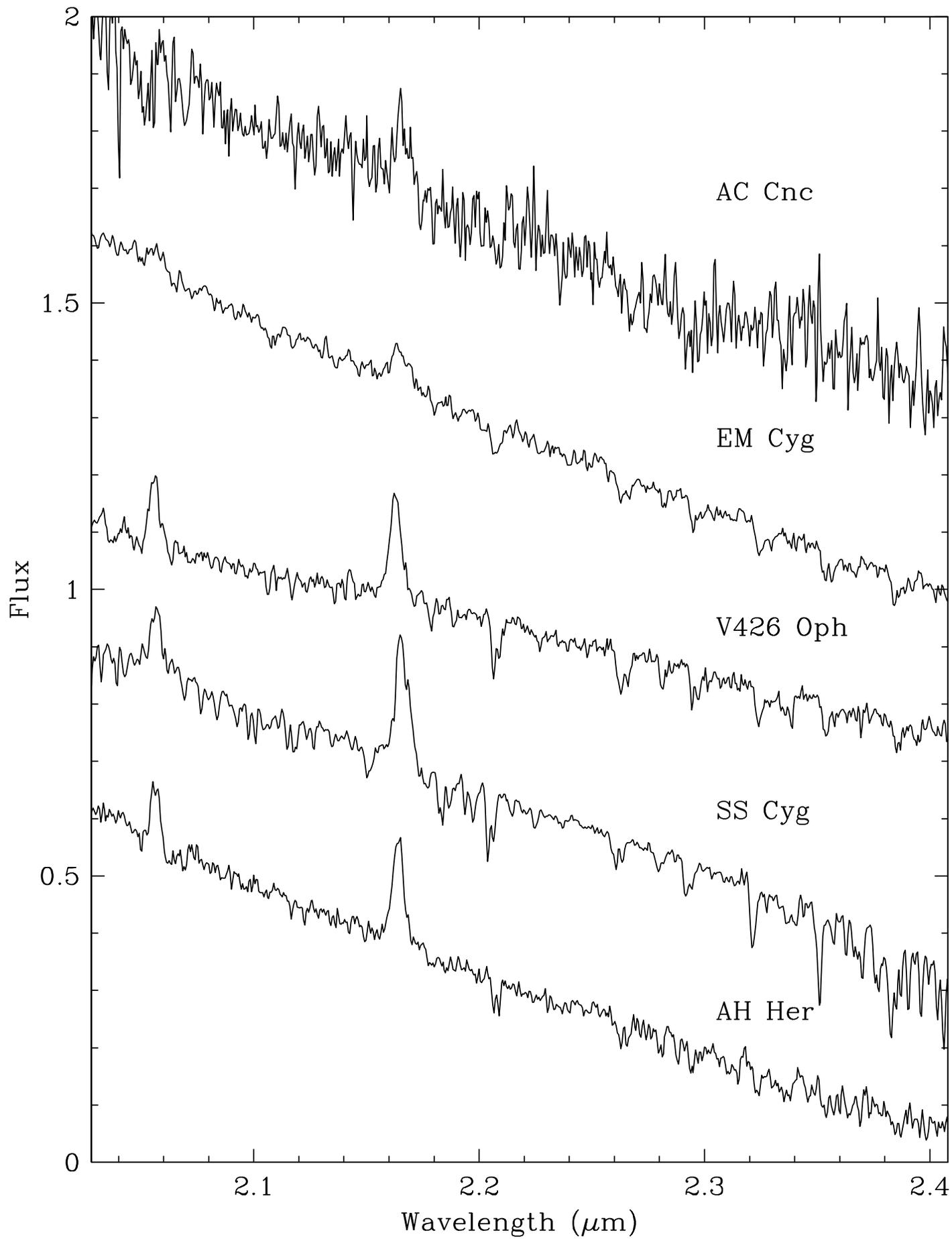
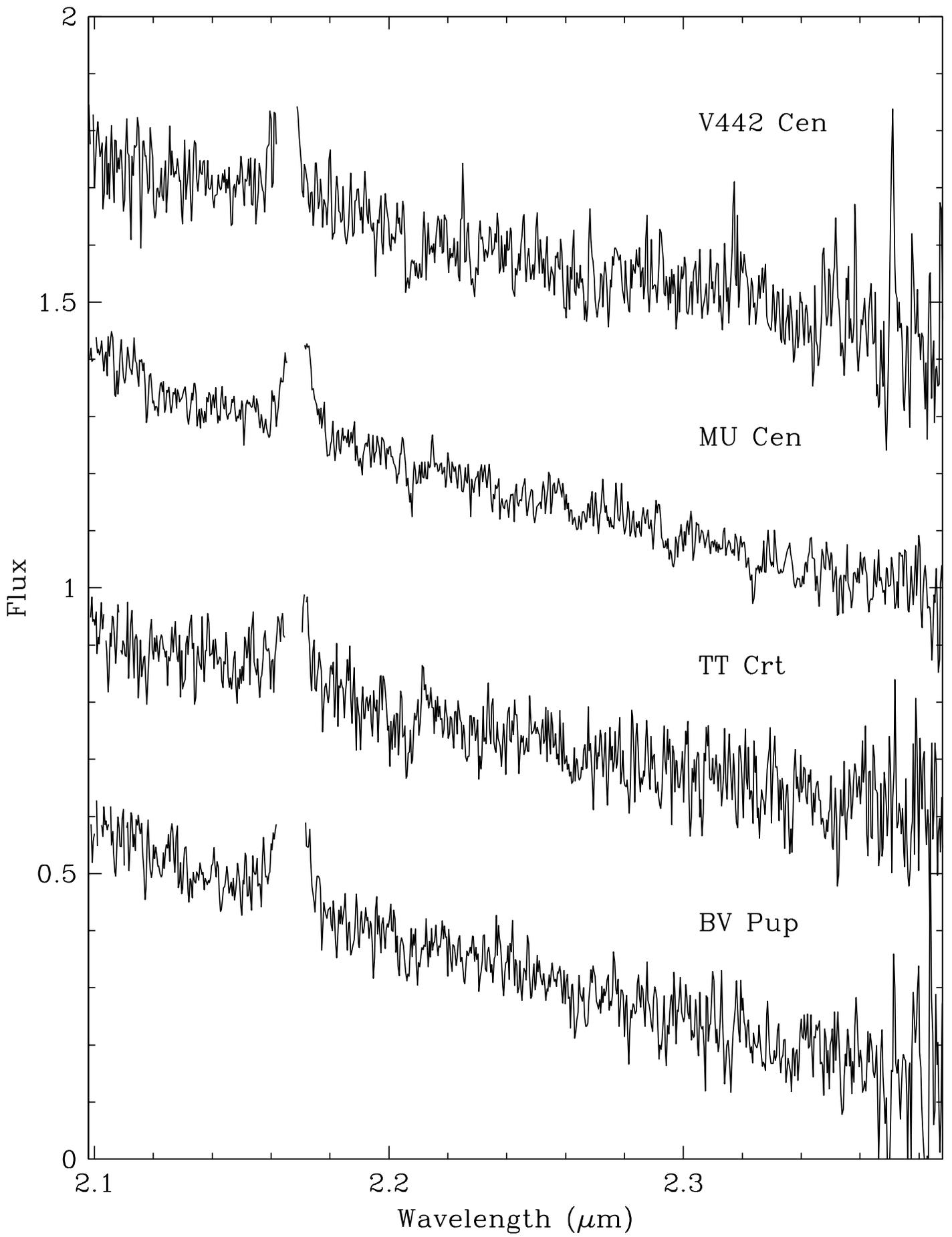
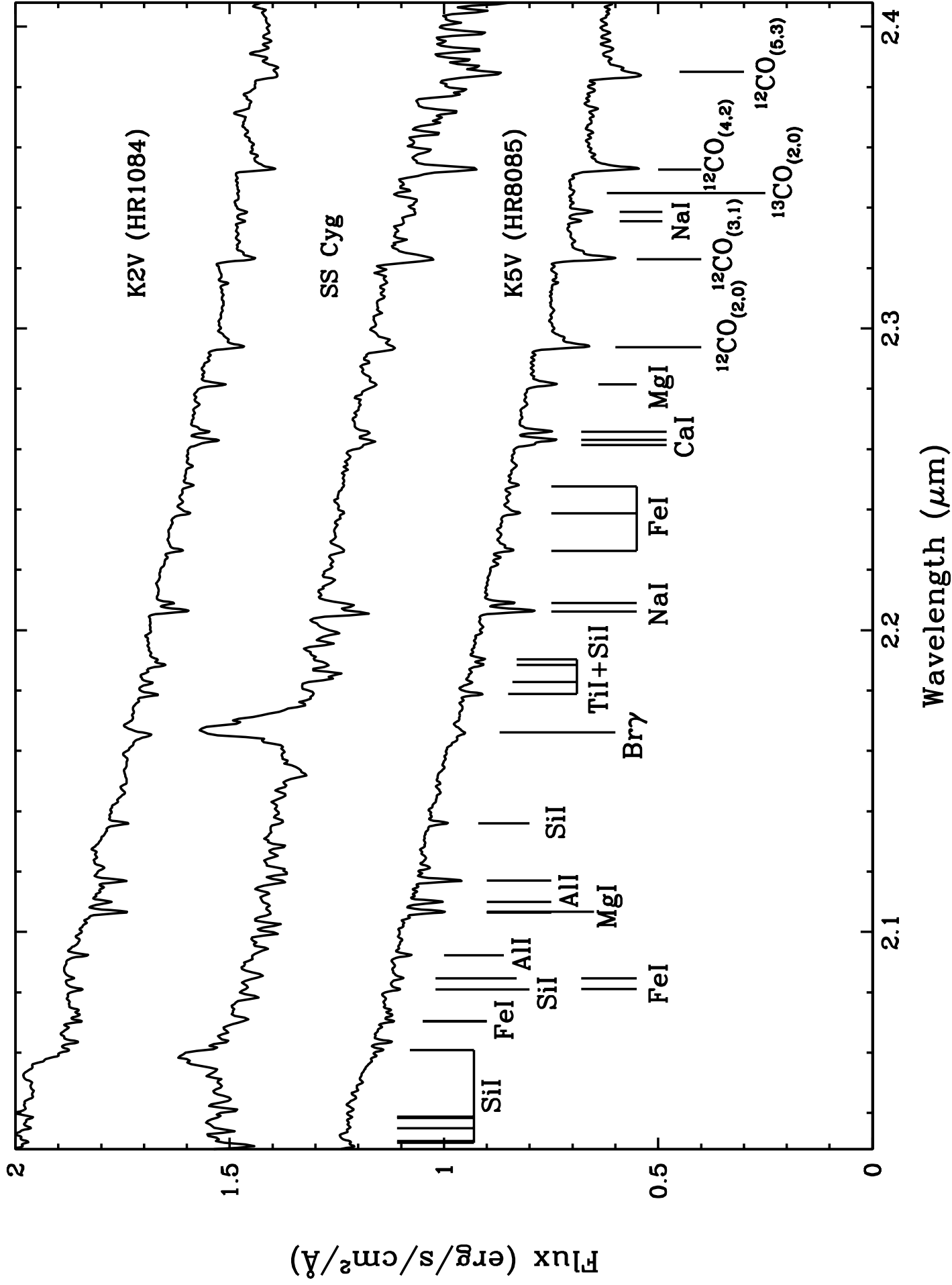
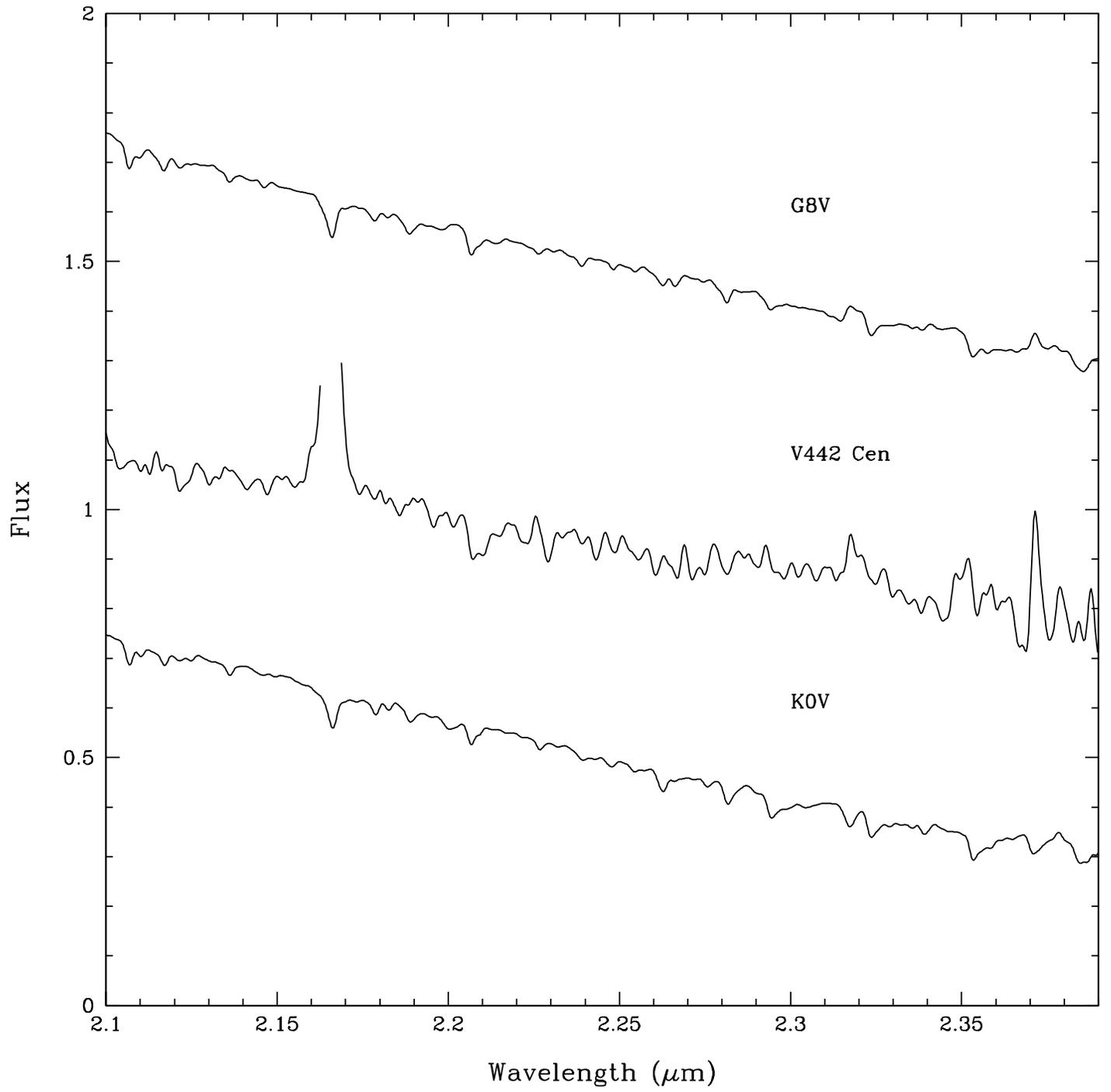
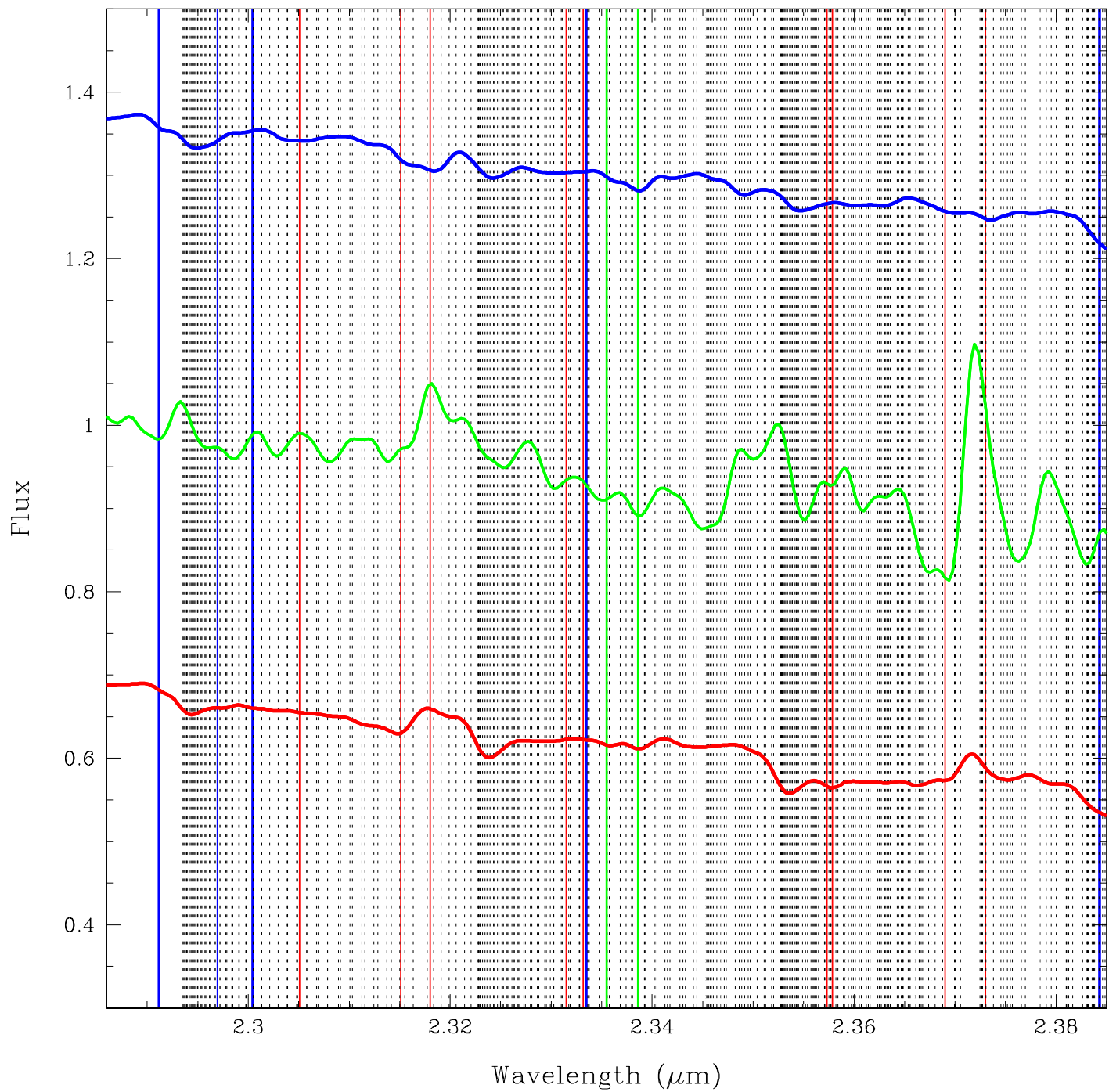
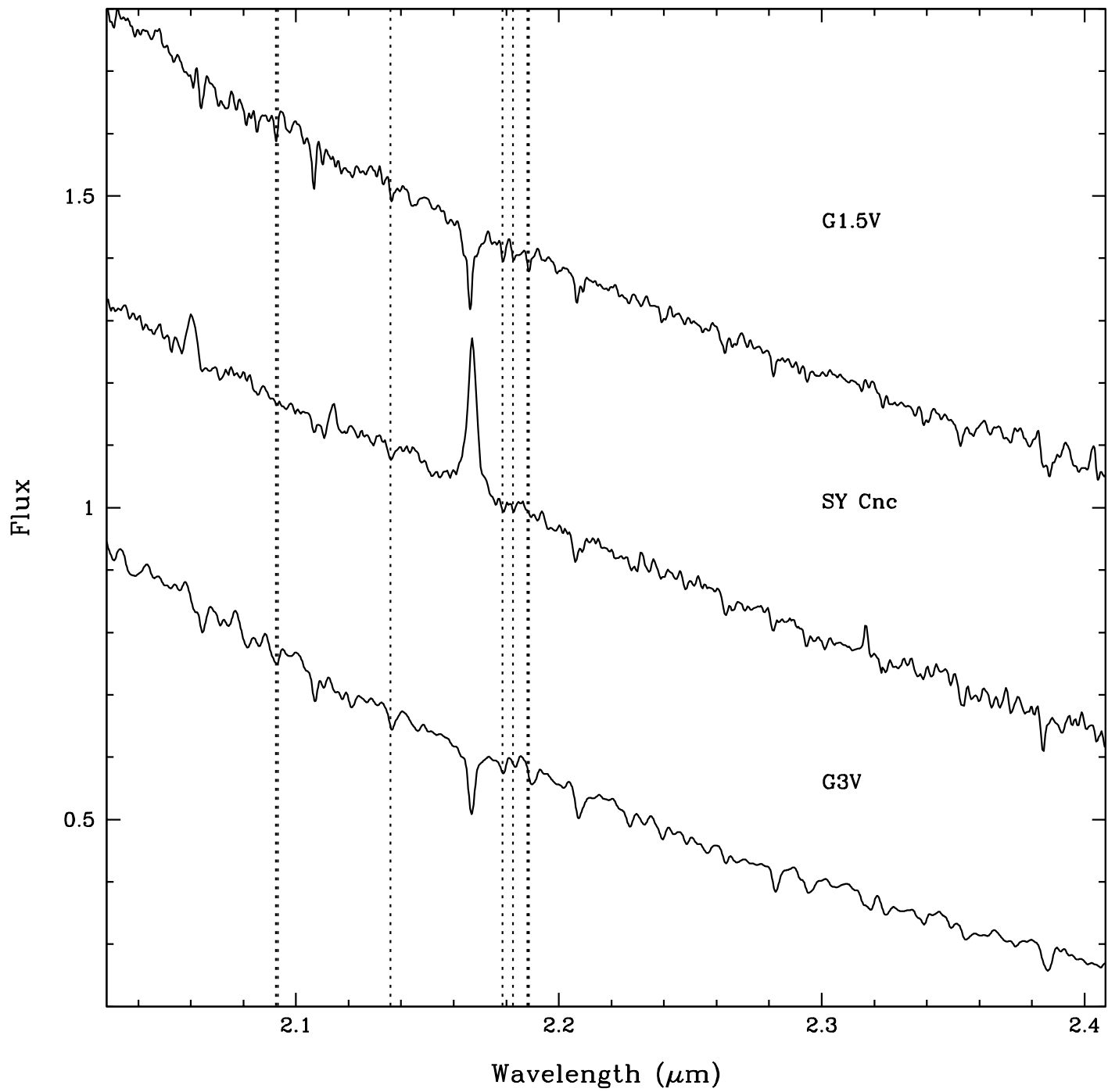
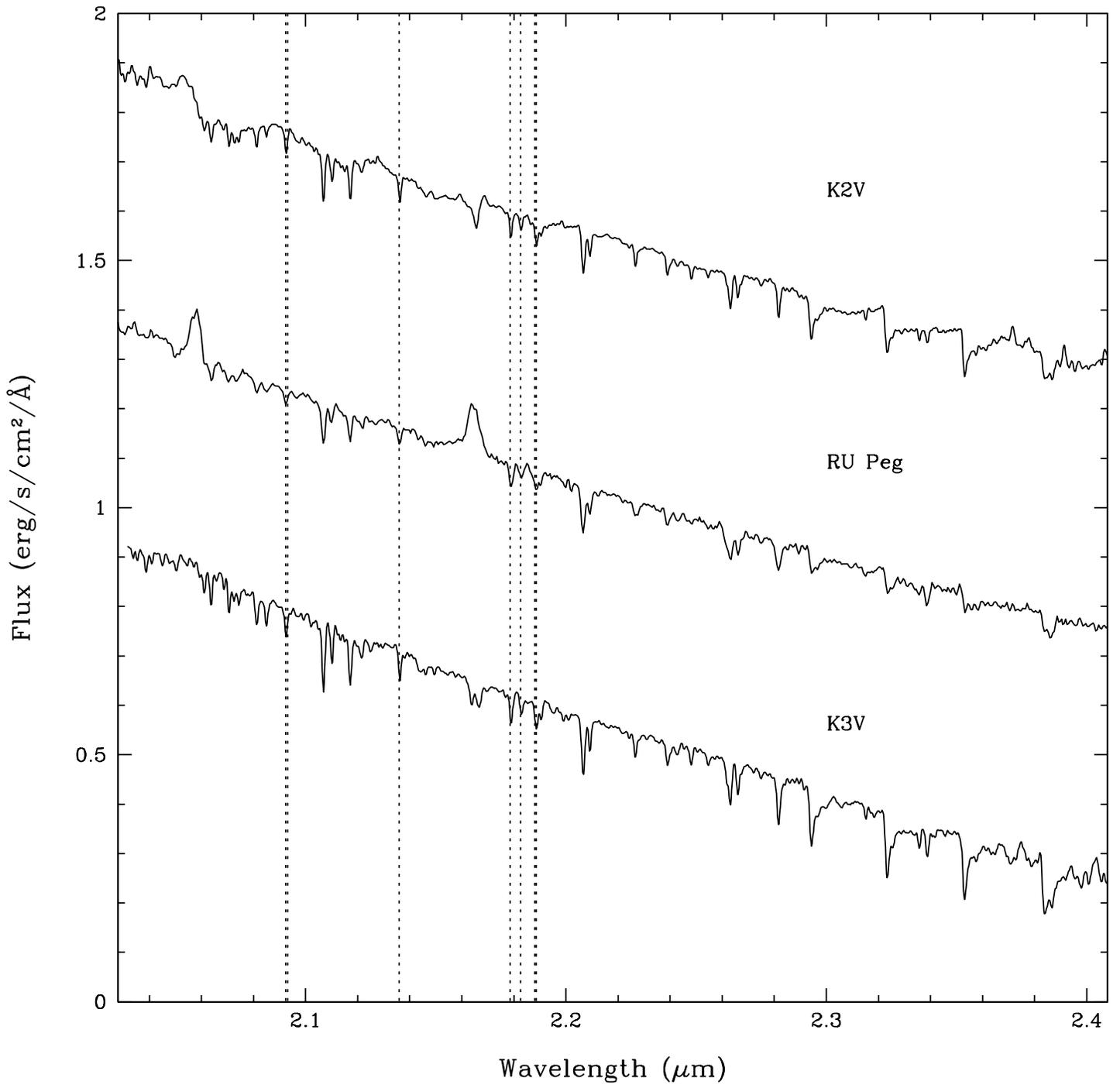
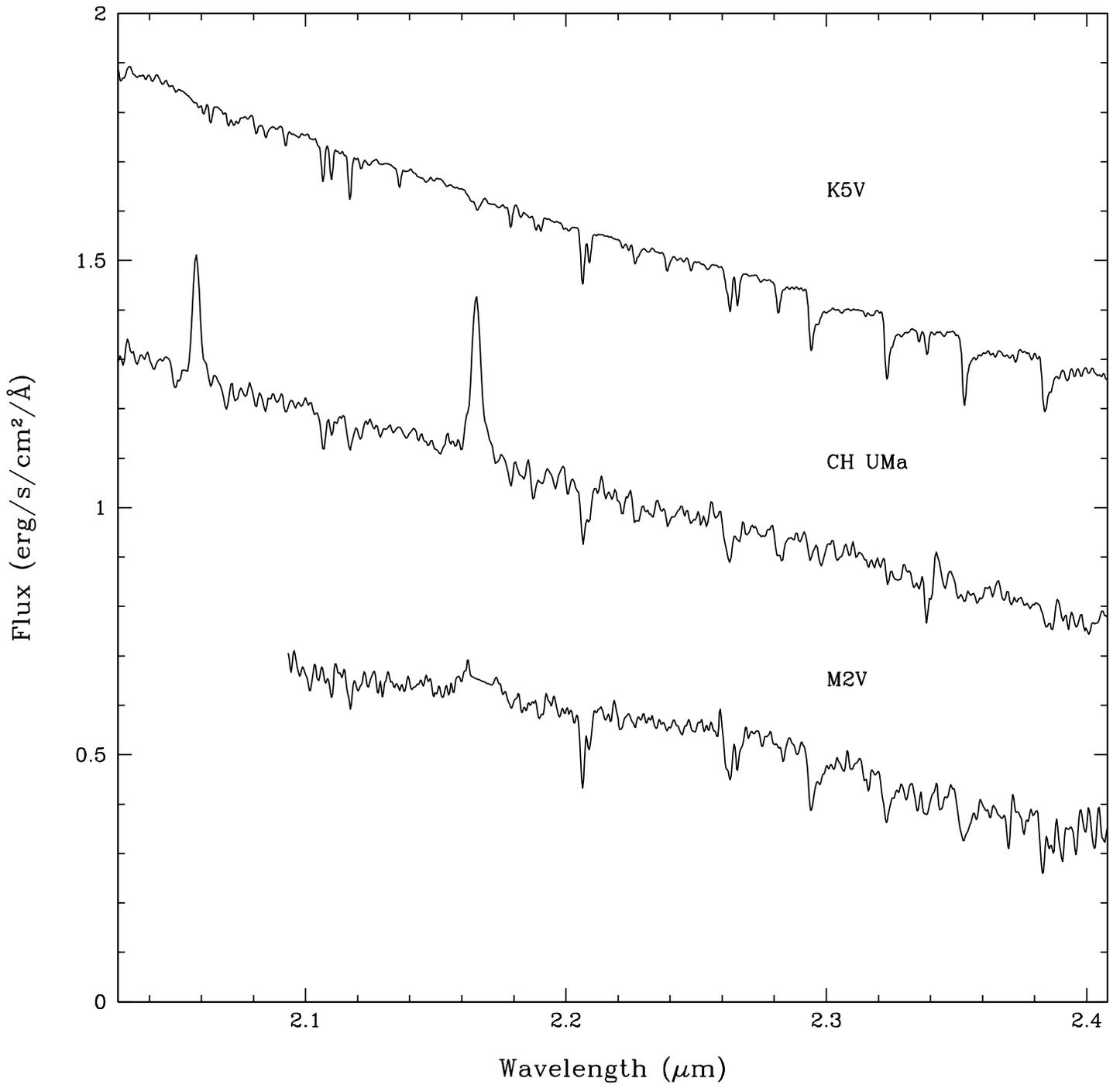
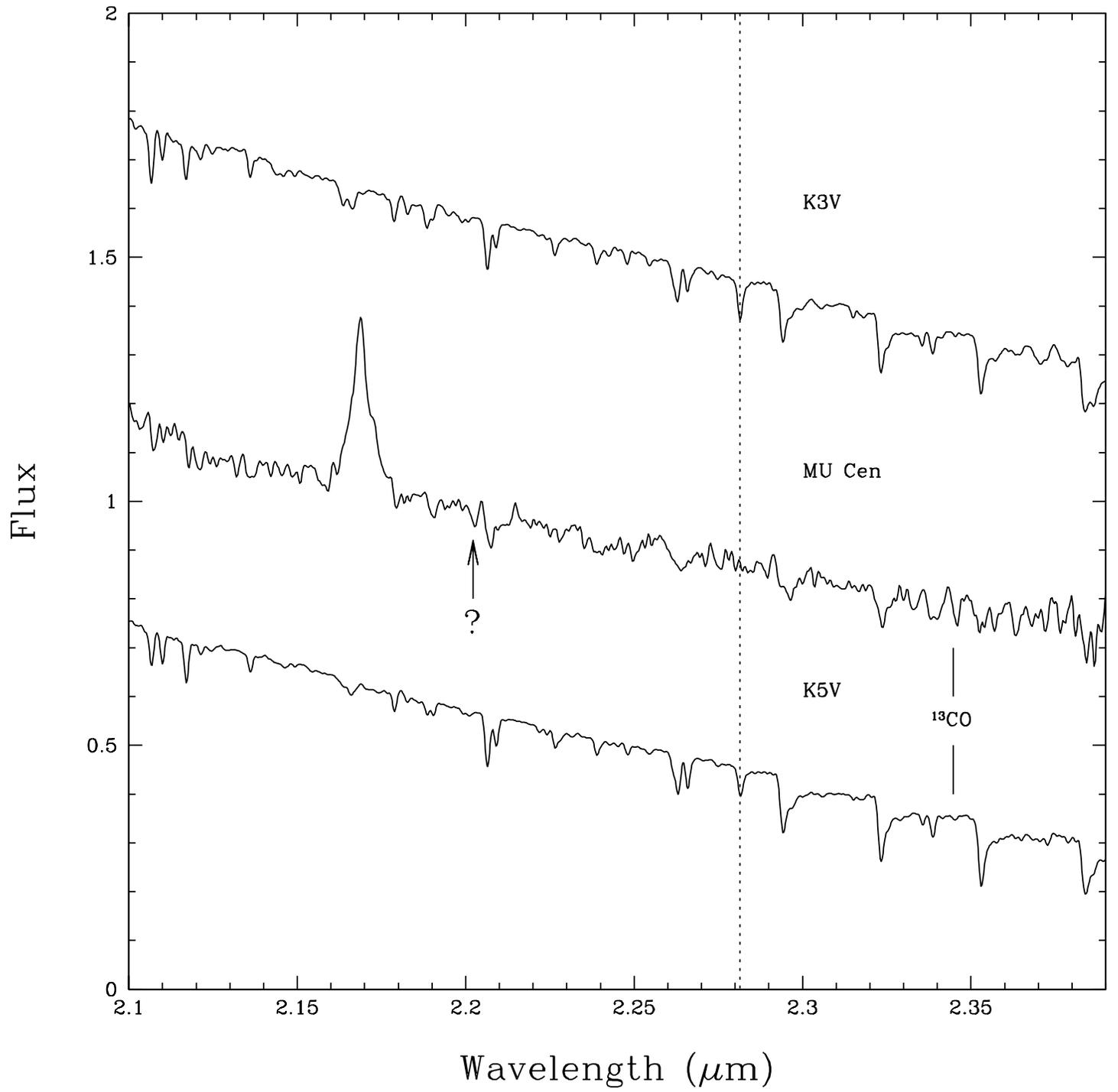
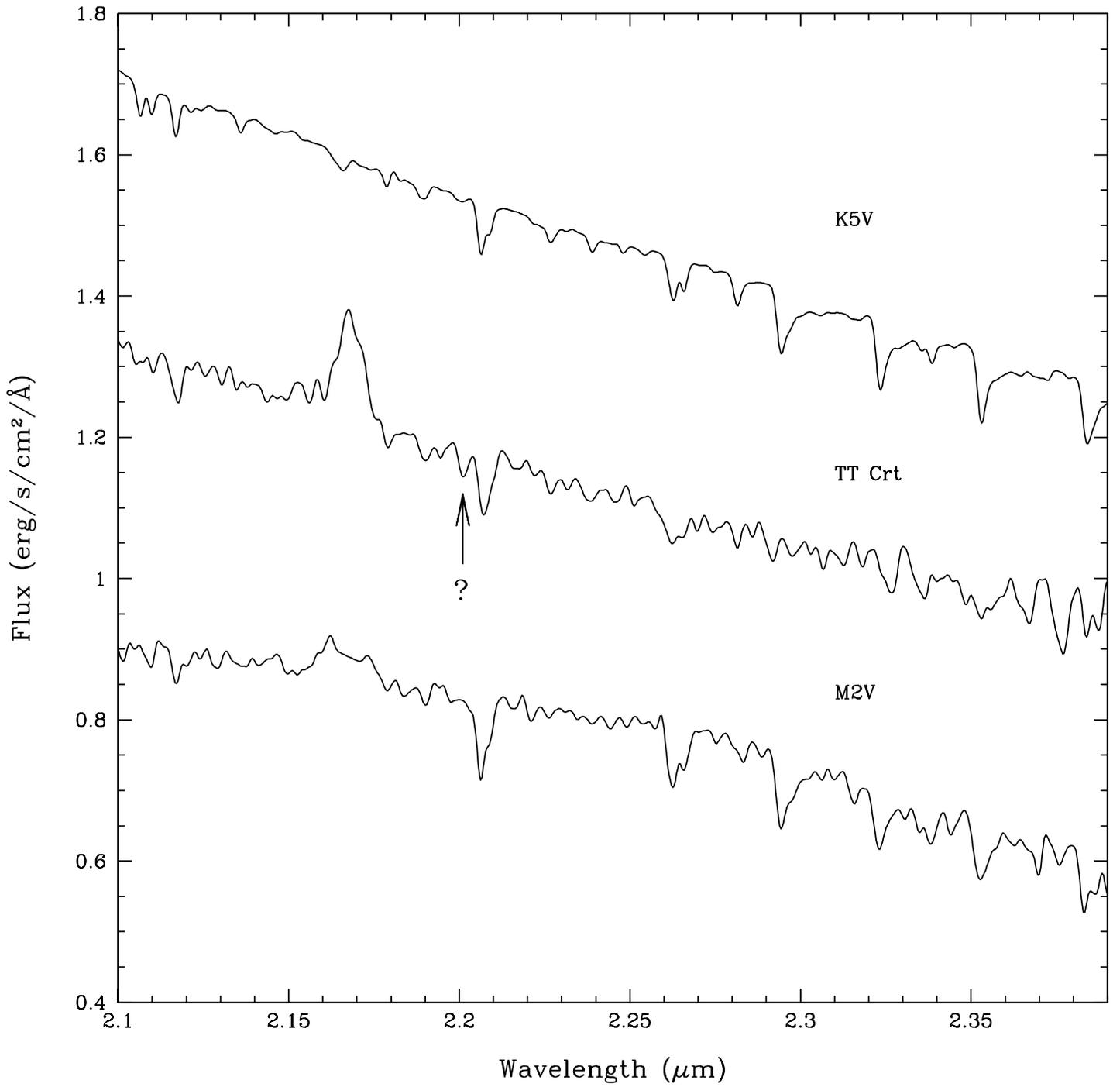
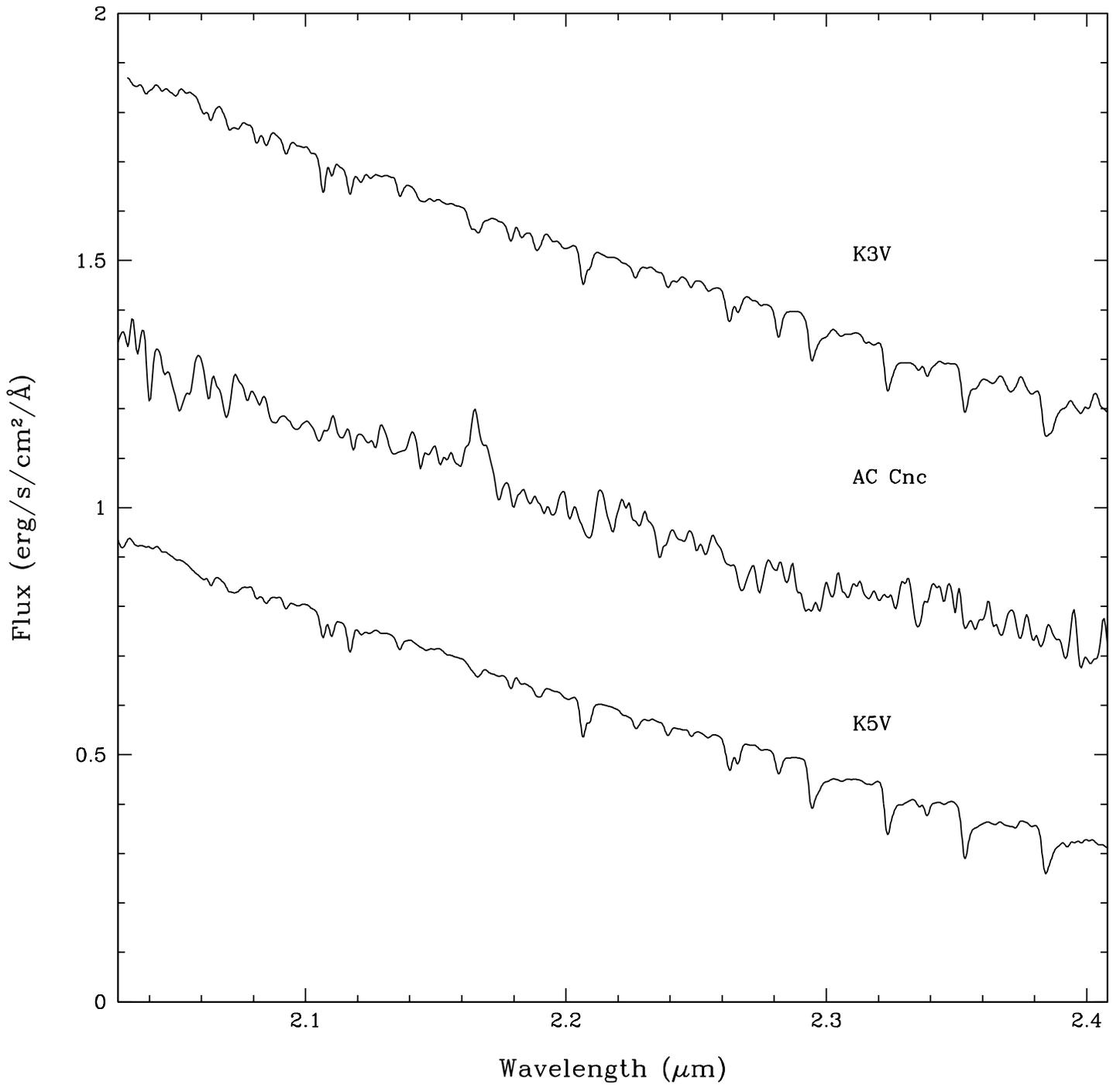
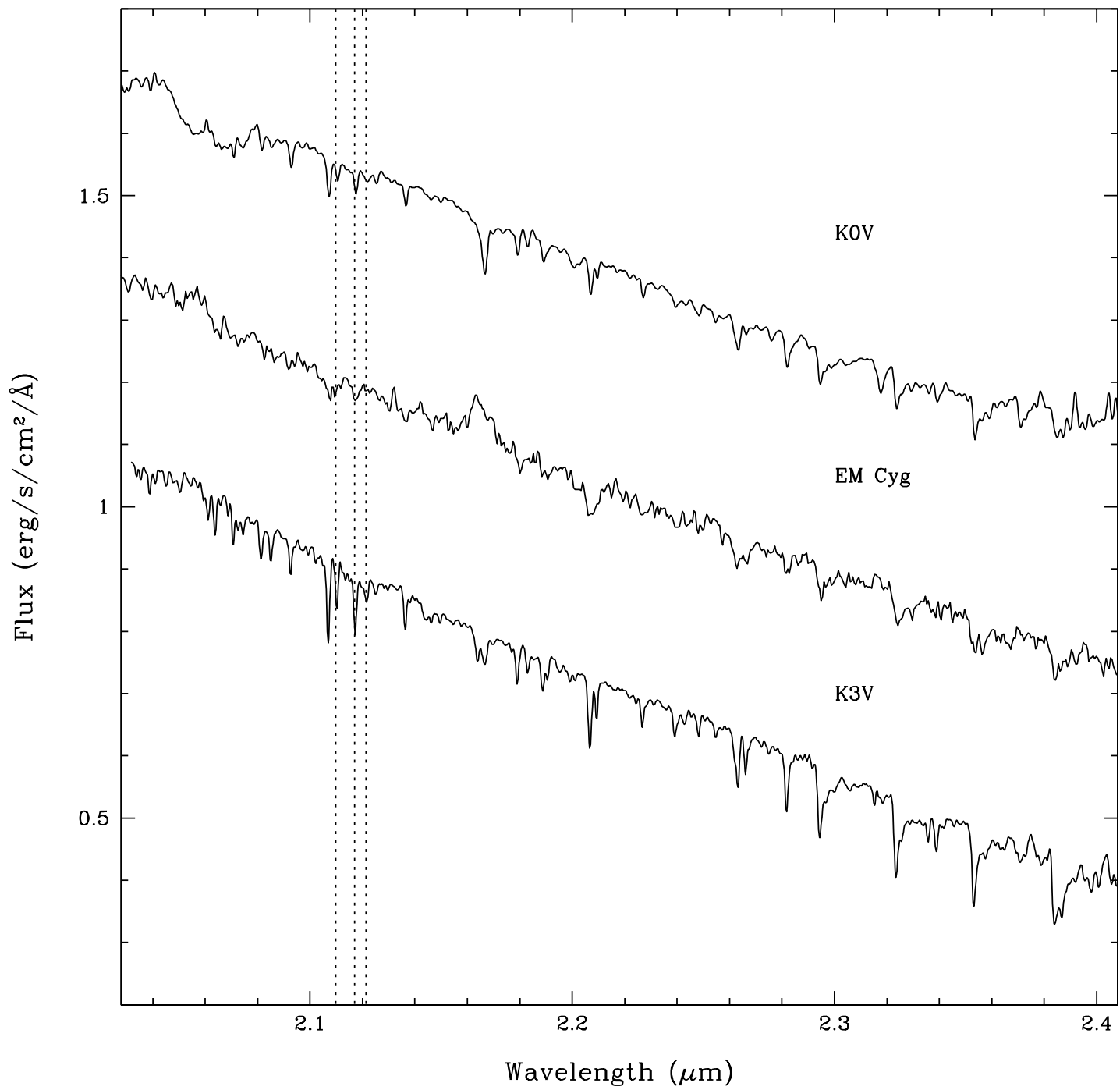
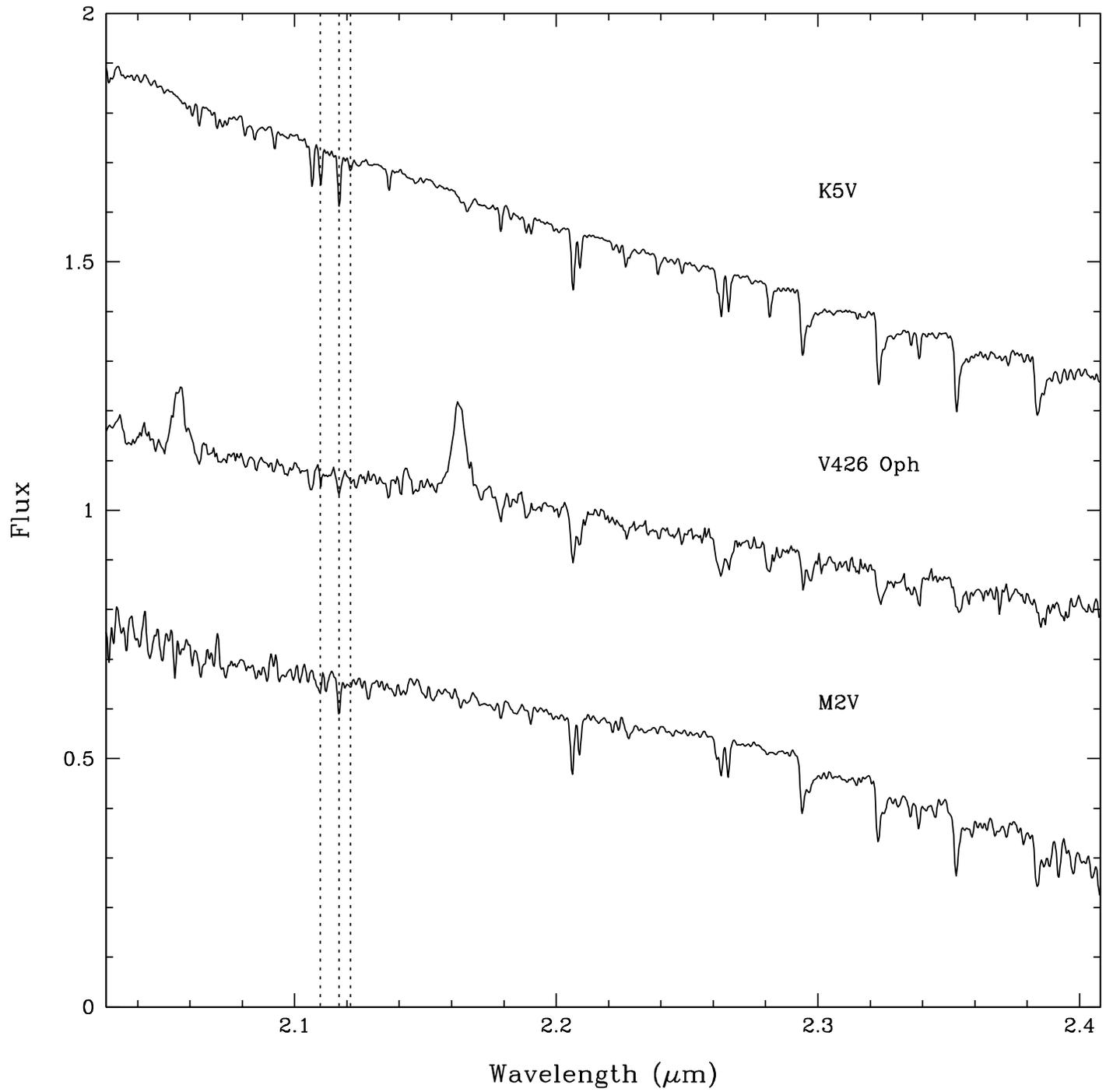
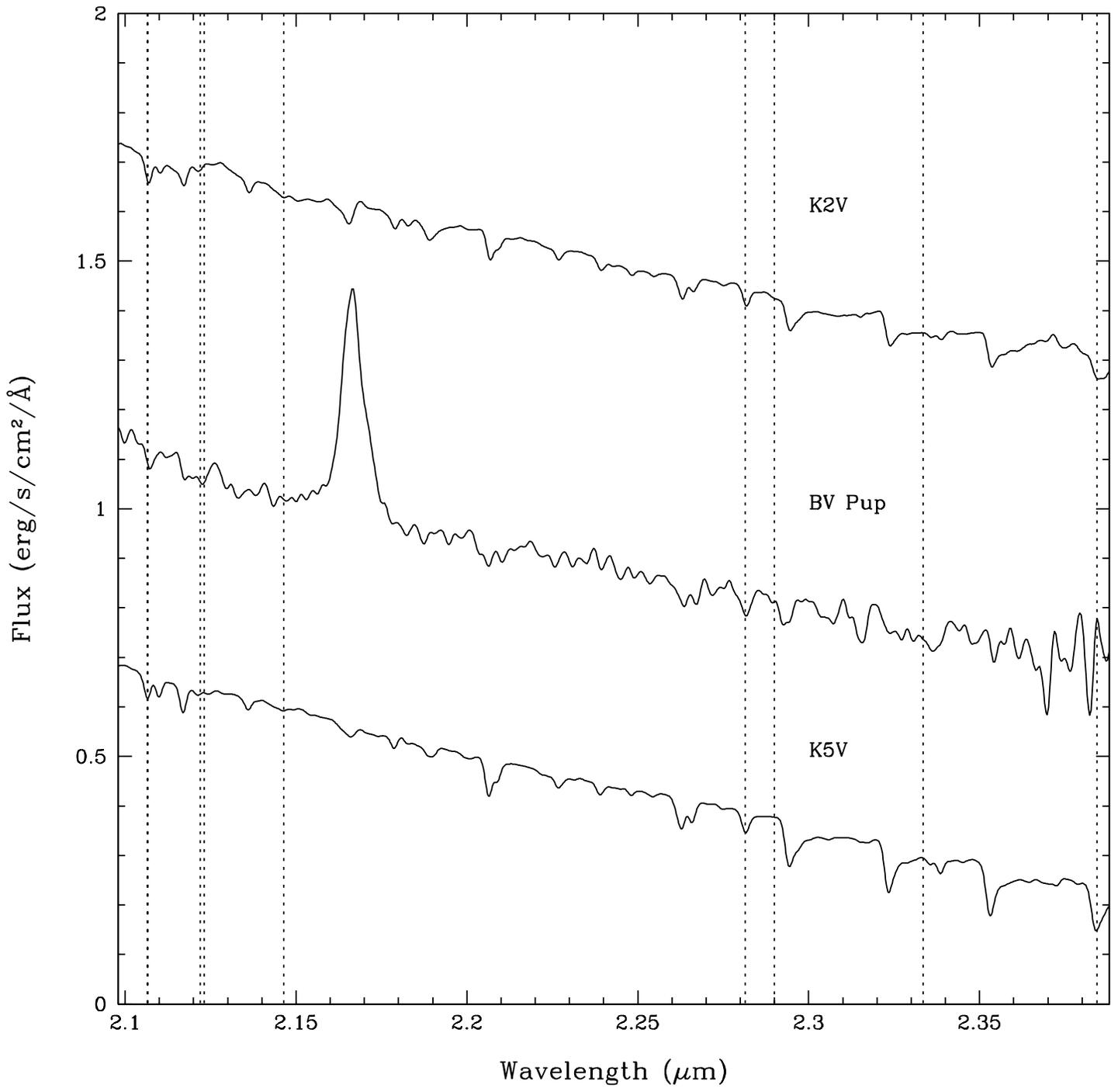
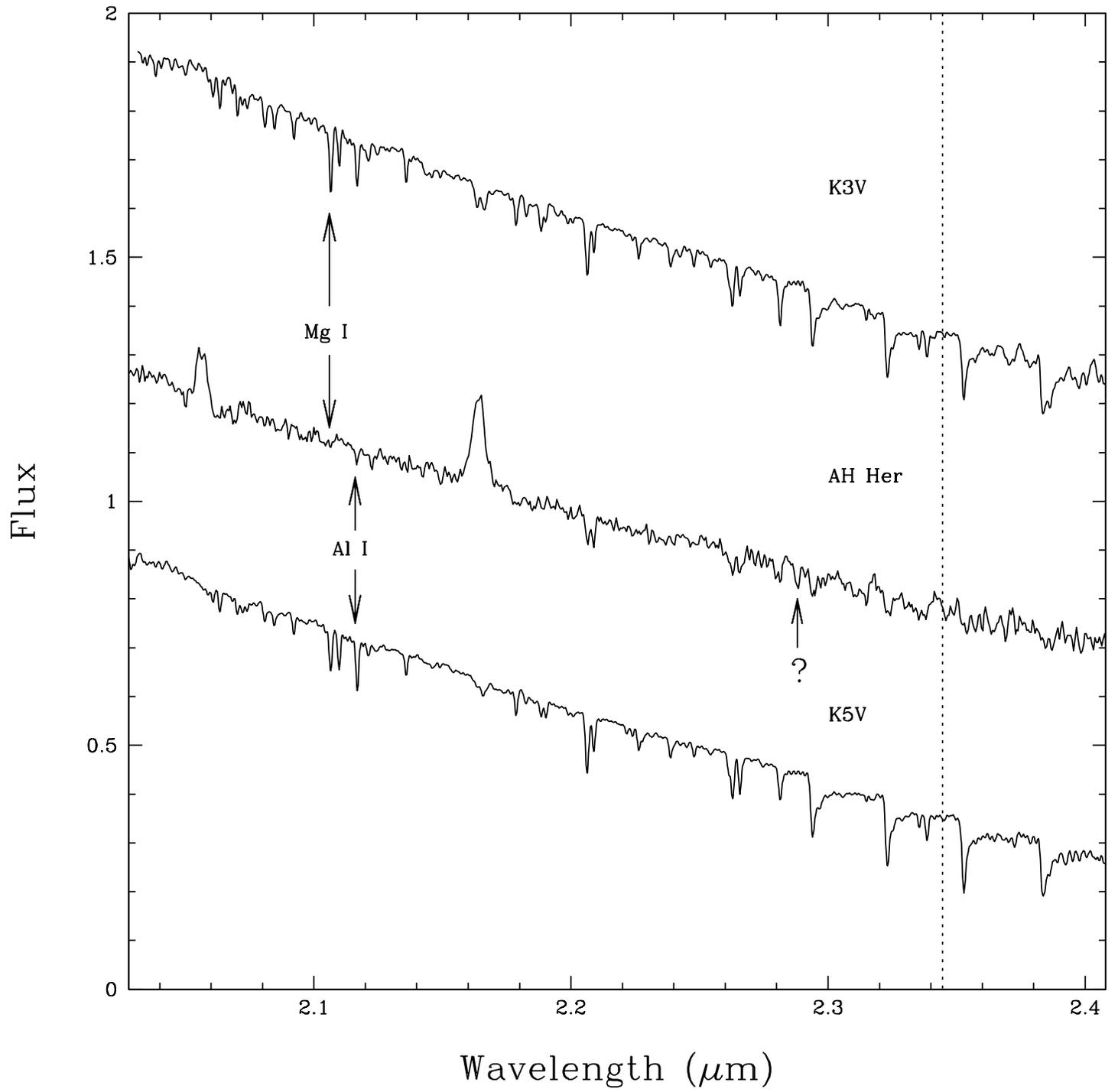
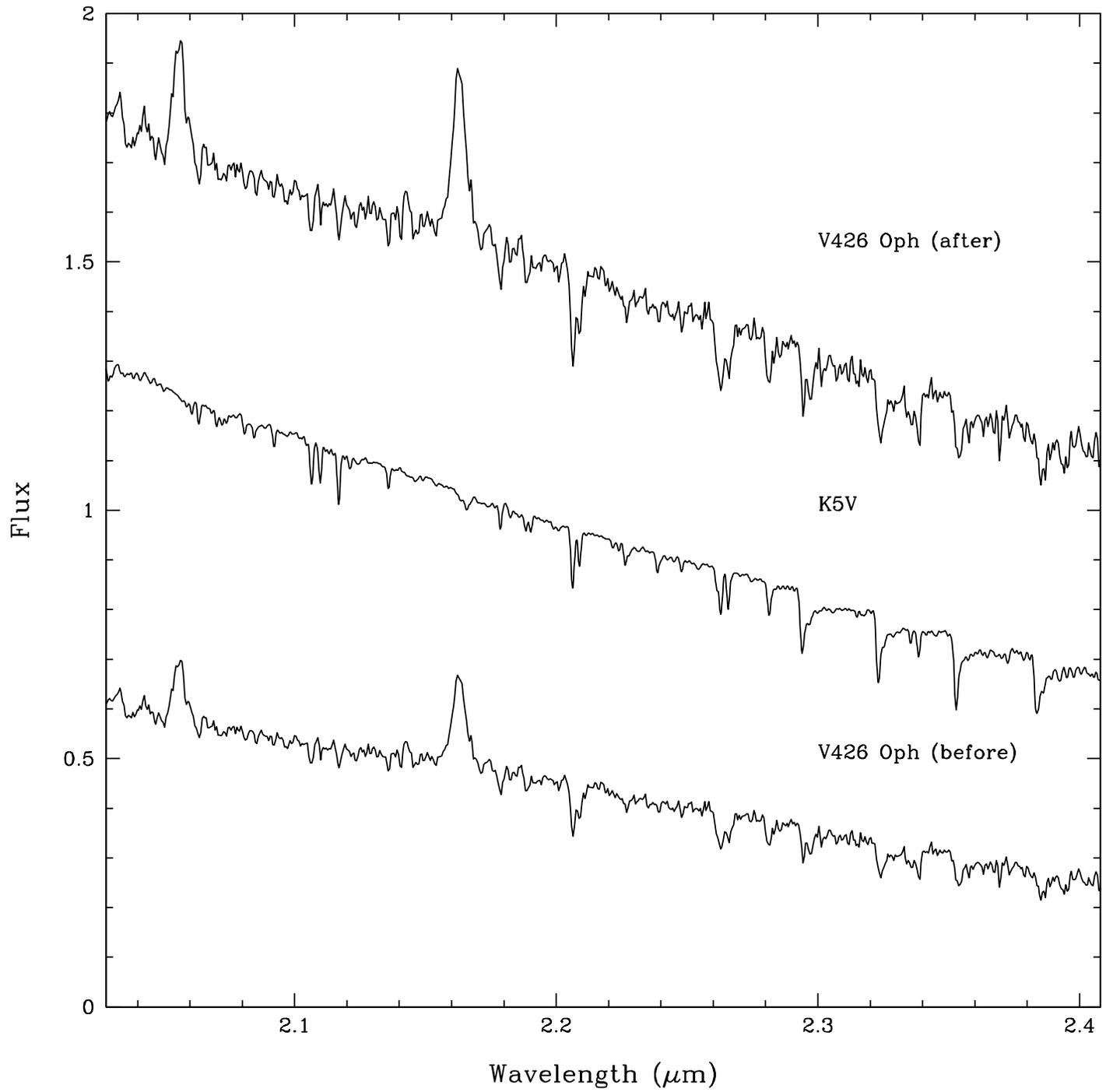
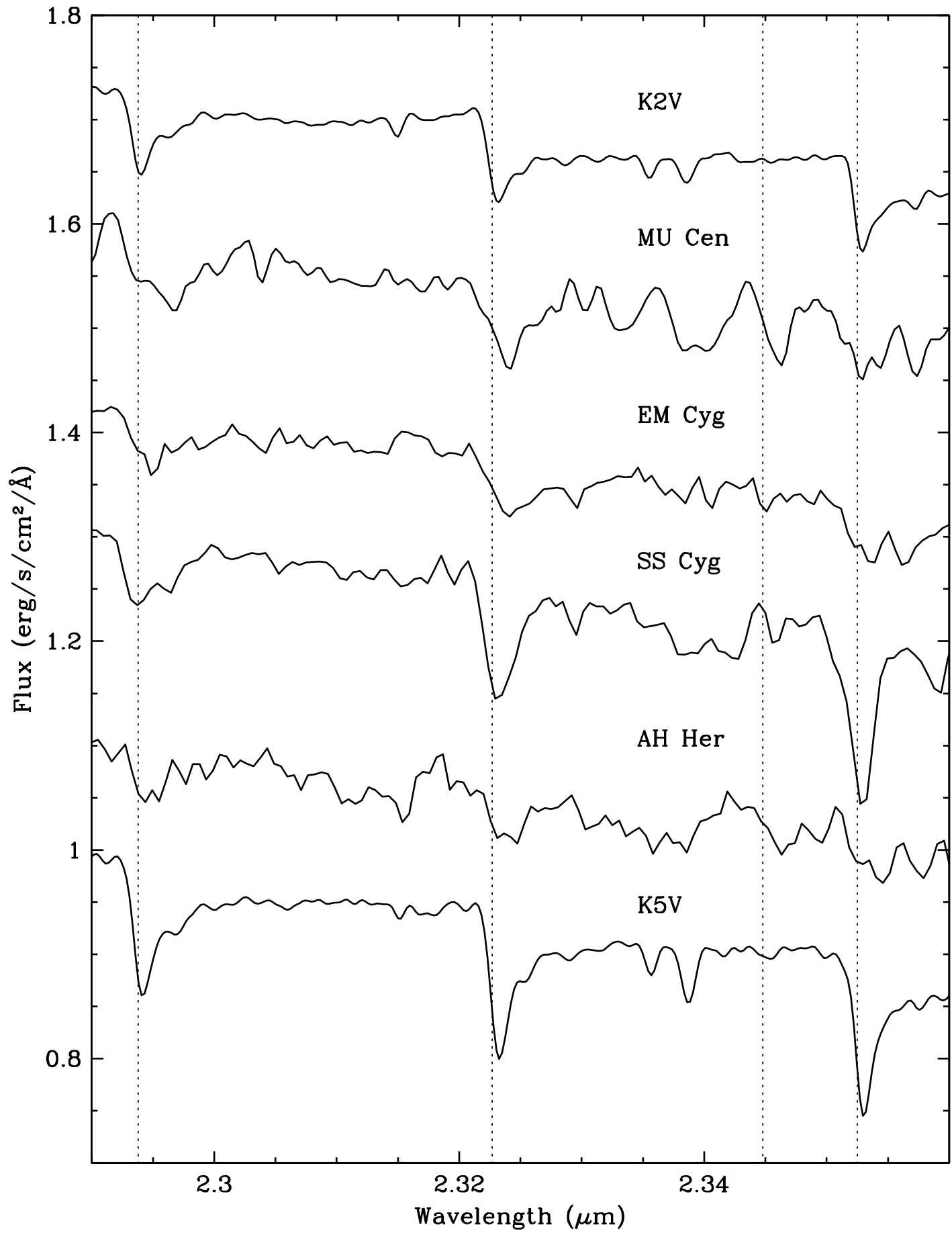
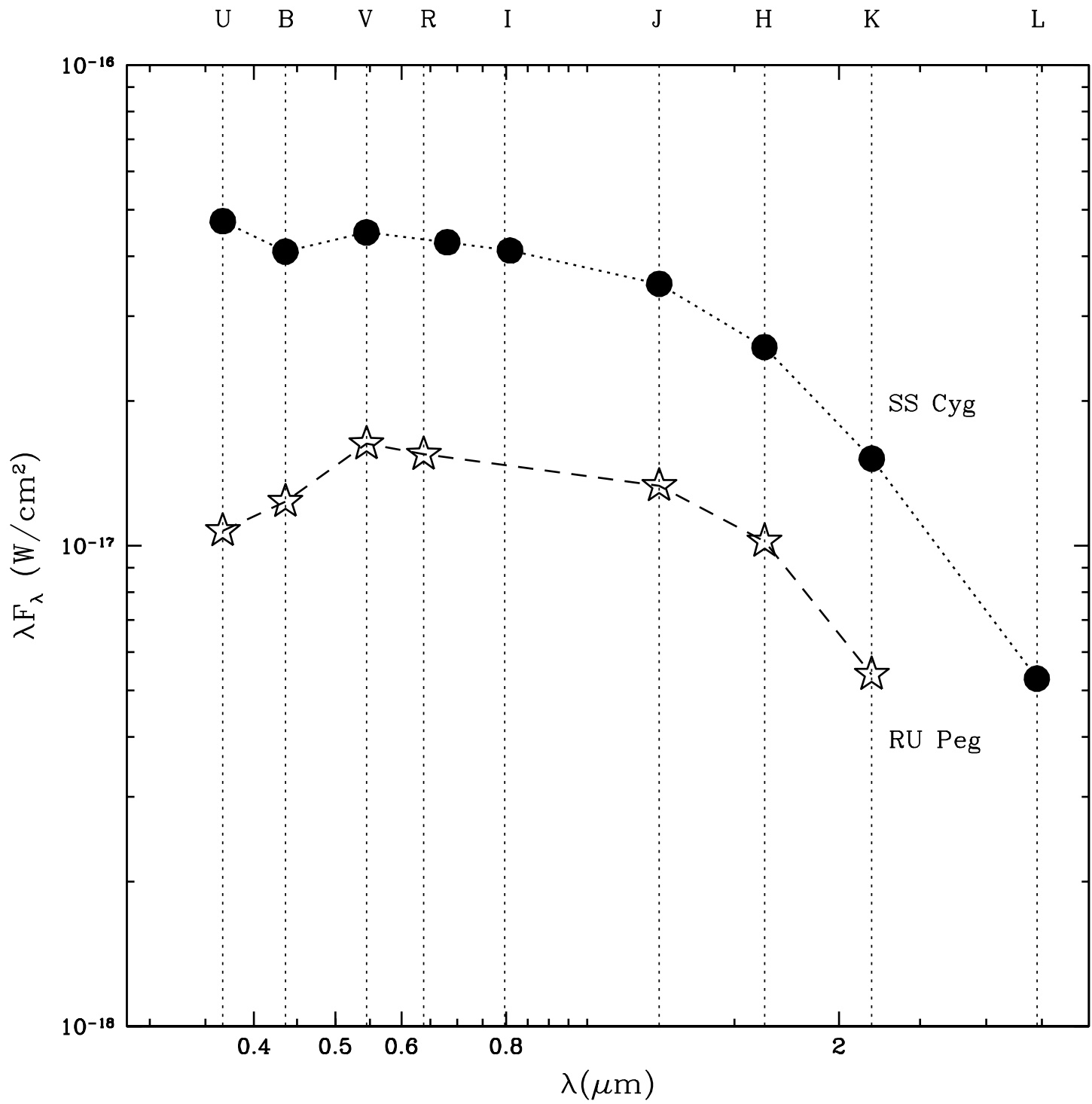
\end{document}